\documentclass[useAMS,usenatbib,fleqn]{mn2e}

\usepackage{amssymb}
\usepackage{times}
\usepackage{graphicx}
\usepackage{epstopdf}
\usepackage{subfigure}
\usepackage[T1]{fontenc}
\usepackage{aecompl} 
\usepackage{multirow}
\usepackage{pdflscape}
\usepackage{rotating}

\usepackage[fleqn]{amsmath}
\usepackage{mathrsfs}
\usepackage{xcolor}

\usepackage{anyfontsize}

\newcommand{\beq}{\begin{equation}}
\newcommand{\eeq}{\end{equation}}

\def\gs{\mathrel{\lower0.6ex\hbox{$\buildrel {\textstyle >}\over{\scriptstyle \sim}$}}}
\def\ls{\mathrel{\lower0.6ex\hbox{$\buildrel {\textstyle <}\over{\scriptstyle \sim}$}}}
\newcommand{\simgt}{\lower.5ex\hbox{$\; \buildrel > \over \sim \;$}}
\newcommand{\simlt}{\lower.5ex\hbox{$\; \buildrel < \over \sim \;$}}

\usepackage{hyperref}

\newcommand{\aap}{A\&A}
\newcommand{\araa}{ARA\&A}
\newcommand{\apj}{ApJ}

\newcommand{\apjs}{ApJS}

\newcommand{\jcap}{J. Cosmol. Astropart. Phys.}

\newcommand{\mnras}{MNRAS}

\newcommand{\ssr}{Space Science Reviews}

\defcitealias{se+et15_comalit_I}{CoMaLit-I}
\defcitealias{ser+al15_comalit_II}{CoMaLit-II}
\defcitealias{ser15_comalit_III}{CoMaLit-III}
\defcitealias{se+et15_comalit_IV}{CoMaLit-IV}
\defcitealias{xxl_I_pie+al16}{XXL-I}
\defcitealias{xxl_II_pac+al15}{XXL-II}
\defcitealias{xxl_III_gil+al15}{XXL-III}
\defcitealias{xxl_IV_lie+al15}{XXL-IV}
\defcitealias{xxl_XIII_eck+al15}{XXL-XIII}

\begin{document}

\title[CoMaLit -- V]{CoMaLit -- V. Mass forecasting with proxies. Method and application to weak lensing calibrated samples.}
\author[M. Sereno \& S. Ettori]{
Mauro Sereno$^{1,2}$\thanks{E-mail: mauro.sereno@unibo.it (MS)}, Stefano Ettori$^{1,3}$
\\
$^1$INAF, Osservatorio Astronomico di Bologna, via Ranzani 1, I-40127 Bologna, Italy\\
$^2$Dipartimento di Fisica e Astronomia, Alma Mater Studiorum -- Universit\`a di Bologna, viale Berti Pichat 6/2, I-40127 Bologna, Italy\\
$^3$INFN, Sezione di Bologna, viale Berti Pichat 6/2, I-40127 Bologna, Italy\\
}


\maketitle

\begin{abstract}
Mass measurements of astronomical objects are most wanted but still elusive. We need them to trace the formation and evolution of cosmic structure but we can get direct measurements only for a minority. This lack can be circumvented with a proxy and a scaling relation. The twofold goal of estimating the unbiased relation and finding the right proxy value to plug in can be hampered by systematics, selection effects, Eddington/Malmquist biases and time evolution. We present a Bayesian hierarchical method which deals with these issues. Masses to be predicted are treated as missing data in the regression and are estimated together with the scaling parameters. The calibration subsample with measured masses does not need to be representative of the full sample  as far as it follows the same scaling relation. We apply the method to forecast weak lensing calibrated masses of the {\it Planck}, redMaPPer and MCXC clusters. {\it Planck} masses are biased low with respect to weak lensing calibrated masses, with a bias more pronounced for high redshift clusters. MCXC masses are under-estimated by $\sim 20$ per cent, which may be ascribed to hydrostatic bias. Packages and catalogs are made available with the paper.
\end{abstract}


\section{Introduction}
\label{sec_intr}

The accurate measurement of the mass of galaxy clusters is crucial to astrophysical investigation \citep{voi05,lim+al13}. Clusters of galaxies have a special place in the hierarchical scenario of structure formation in a universe dominated early by cold dark matter and later by dark energy in the form of a cosmological constant. They are the most massive and latest objects to near viral equilibrium. 

Theoretical models and numerical simulations of formation and evolution of cosmic structure are naturally expressed in terms of mass. Still, precise mass measurements are elusive and very difficult to get in the large samples detected by ongoing and future surveys \citep{eucl_lau_11,xxl_I_pie+al16}. The weak lensing (WL) mass, the X-ray mass relying on hydrostatic equilibrium (HE) or the mass based on galaxy dynamics can provide accurate estimates but they can be either very expansive to get or reliable only for regular objects. WL and caustics masses are in principle independent of the equilibrium state of the clusters but are affected by their own systematics \citep{gar+al04,men+al10,be+kr11,ras+al12,sve+al15} and up to date are available only in small samples \citep{wtg_III_14, ume+al16b,hoe+al15,ok+sm16,rin+al13}.

The challenge is then to predict the mass given a cluster property (the proxy) which can be easily measured and that is strictly related to the mass, e.g. optical richness, X-ray luminosity, Sunyaev-Zel'dovich (SZ) flux. This requires the accurate calibration of the observable proxy through comparison with direct mass estimates \citep{an+be12,ett13}. 

The main properties of galaxy clusters are expected to scale with the mass. If gravity is the dominant process, clusters are self-similar and scaling relations in form of scale-free power laws subsist among cluster properties \citep{kai86,gio+al13}. This picture is substantially confirmed by observations \citep{ett13,ett15} and numerical simulations \citep{sta+al10,fab+al11,ang+al12}, which showed that intrinsic scatters around the median relations approximately follow a log-normal distribution. Deviations from the self-similar scheme come from non-gravitational processes, e.g. feedback and non-thermal processes, which can contribute significantly to the global energy budget \citep{mau+al12}.

The best case scenario for a proxy is an observable which can be measured quickly and accurately and that can be given as input to a calibrating function whose output is the mass. This is not the scenario we live in. It is rare to measure the proxy with a level of accuracy such that we can consider the measurement uncertainty negligible. Furthermore, selection biases can make the measured proxy not representative of the nominal value we should use in the scaling relation. Finally, the knowledge of the scaling relation is uncertain too.

Scaling relation relations are not deterministic. We rather estimate the probability of a mass given a proxy. The width of the probability density gives rise to an intrinsic scatter which limits the accuracy to which the mass can be estimated. Furthermore, the scaling relation is known with its own uncertainties. Significant efforts in the field of Bayesian hierarchical analysis have been pursued to efficiently derive scaling relations \citep{dag05,kel07,an+be12,mau14,man16,ser16_lira}.

This is the fifth in the CoMaLit (COmparing MAsses in LITerature) series of papers, wherein we have been applying Bayesian hierarchical methods with latent variables to comparative studies of masses and scaling relations. In the first paper of the series \citep[ CoMaLit-I]{se+et15_comalit_I}, we considered the best rout to calibrate scaling relations. We looked for systematic differences in WL and HE masses and we assessed the overall level of the intrinsic scatters. In the second paper of the series \citep[ CoMaLit-II]{ser+al15_comalit_II}, we dealt with the measurements of scaling relations. We presented the Bayesian method and we applied it to WL clusters with measured SZ flux. The third paper of the series \citep[ CoMaLit-III]{ser15_comalit_III} introduced the Literature Catalogs of weak Lensing Clusters of galaxies (LC$^2$), which we have been using to calibrate mass proxies. The fourth paper of the series \citep[ CoMaLit-IV]{se+et15_comalit_IV} dealt with time-evolution and completeness. The scalings of optical richness, X-ray luminosity and galaxy velocity dispersion with mass were considered. Here, we deal with efficient mass prediction. 

The present paper is structured as follows. In Sec.~\ref{sec_expe}, we discuss the basics of a mass proxy. Sources of uncertainty, systematics and biases in mass prediction are listed in Sec.~\ref{sec_delta}. The Bayesian hierarchical model is presented in Sec.~\ref{sec_pred}. The calibration sample is introduced in Sec.~\ref{sec_samp}. Mass forecast of {\it Planck}, redMaPPer or MCXC clusters is presented in Secs.~\ref{sec_planck}, \ref{sec_redmapper} and \ref{sec_mcxc}, respectively. The comparison of weak lensing calibrated masses is presented in Sec.~\ref{sec_MWLc_comp}. In Sec.~\ref{sec_reli}, we discuss the reliability of heterogeneous calibration samples. Final considerations are in Sec.~\ref{sec_conc}. Appendix~\ref{app_imple} details the  implementation of the forecasting.

\subsection{Conventions and notations}

Throughout the series of papers, the frame-work cosmological model is the concordance flat $\Lambda$CDM universe with matter density parameter $\Omega_\text{M}=1-\Omega_\Lambda=0.3$ and Hubble constant $H_0=70~\mathrm{km~s}^{-1}\mathrm{Mpc}^{-1}$; $H(z)$ is the redshift dependent Hubble parameter and $E_z\equiv H(z)/H_0$.  

The redshift dependence $T$ is expressed as $F_z$, which can be proportional to either $E_z$ or $(1+z)$. $D$ is either the luminosity or the angular diameter distance. The reference redshift is denoted as $z_\text{ref}$. $F_z$ and $D$ are normalized such that $F_z(z_\text{ref})=1$ and $D(z_\text{ref})=1$.
In this paper we set $F_z = E_z/E_z(z_\text{ref})$ with $z_\text{ref}=0.01$.

$O_{\Delta}$ denotes a global property of the cluster measured within the radius which encloses a mean over-density of $\Delta$ times the critical density at the cluster redshift, $\rho_\mathrm{cr}=3H(z)^2/(8\pi G)$. 

`$\log$' is the logarithm to base 10 and `$\ln$' is the natural logarithm.

Computations were performed with the \textsc{R}-package \texttt{LIRA}\footnote{The package \texttt{LIRA} (LInear Regression in Astronomy) is publicly available from the Comprehensive R Archive Network at \url{https://cran.r-project.org/web/packages/lira/index.html}. Beta versions can be downloaded from GitHub at \url{https://github.com/msereno/lira}. For further details, see \citet{ser16_lira}}. 
The full list of parameters considered in the most general regression scheme is presented in Table~\ref{tab_par}. The actual implementation for the cases studied in the present paper is detailed in Secs.~\ref{sec_pred}, \ref{sec_planck}, \ref{sec_redmapper} and \ref{sec_mcxc} and in Appendix~\ref{app_imple}.

Products associated with the present paper and the CoMaLit series are hosted at \url{http://pico.bo.astro.it/\textasciitilde sereno/CoMaLit/}.

\begin{table*}
\caption{
List of the parameters of the regression scheme and their description. The variables $Z$ is the covariate, $X$ is a proxy of $Z$, and $Y$ is the response variable. Priors in square brackets can be set only as delta distributions. See the \texttt{LIRA} user manual for details.}
\label{tab_par}
\centering
\resizebox{0.95\hsize}{!} {
\begin{tabular}[c]{l l l l l}
	\hline
	Type & Meaning & Symbol & Code symbol & Default prior\\ 
	\noalign{\smallskip}  
	\hline
	\multicolumn{5}{c}{\it $Y$-$Z$ scaling} \\
	 \noalign{\smallskip}  
	\multicolumn{5}{l}{$Y_Z = \alpha_{Y|Z}+\beta_{Y|Z} Z + \gamma_{Y|Z} T+ \delta_{Y|Z} Z\ T$} \\
	\noalign{\smallskip}  
	Conditional scaling relation &	intercept & $\alpha_{Y|X}$ & \texttt{alpha.YIZ} & \texttt{dunif}\\
	 & slope & $\beta_{Y|X}$  & \texttt{beta.YIZ} & \texttt{dt}\\
	& time evolution & $\gamma_{Y|Z}$  & \texttt{gamma.YIZ} & \texttt{dt}\\
	& time tilt & $\delta_{Y|Z}$  & \texttt{delta.YIZ} & \texttt{0}\\
	\noalign{\smallskip}
	\multicolumn{5}{l}{$Y_Z = \alpha_{Y|Z,\text{knee}}+\beta_{Y|Z,\text{knee}} Z + \gamma_{Y|Z,\text{knee}} T+ \delta_{Y|Z,\text{knee}} Z\ T$} \\
	\noalign{\smallskip}  
	Scaling relation before the break  & slope for $Z<Z_\text{knee}$    & $\beta_{Y|X,\text{knee}}$  & \texttt{beta.YIZ.knee} & \texttt{beta.YIZ}\\
		                                          & time tilt  for $Z<Z_\text{knee}$    & $\delta_{Y|Z,\text{knee}}$  & \texttt{delta.YIZ.knee} & \texttt{delta.YIZ}\\
	\noalign{\smallskip}
	\multicolumn{5}{l}{$f_\text{knee}(Z)=1/(1+\exp [(Z-Z_\text{knee})/l_\text{knee} ])$} \\
	\noalign{\smallskip}  
	Transition function  & break scale    & $Z_\text{knee}$  & \texttt{Z.knee} & \texttt{dunif}\\
		                       & break length   & $l_\text{knee}$   & \texttt{l.knee} & \texttt{1e-04}\\
	\multicolumn{5}{c}{\it $X$-$Z$ scaling} \\
	\noalign{\smallskip}
		\multicolumn{5}{l}{$X_Z = \alpha_{X|Z}+\beta_{X|Z} Z + \gamma_{X|Z} T + \delta_{X|Z} Z\ T$} \\
	\noalign{\smallskip}  
	Proxy of the independent variable &	bias & $\alpha_{X|Z}$ & \texttt{alpha.XIZ} & \texttt{0}\\
	 & slope & $\beta_{X|Z}$  & \texttt{beta.XIZ} & \texttt{1}\\
	& time evolution & $\gamma_{X|Z}$  & \texttt{gamma.XIZ} & \texttt{0}\\
	& time tilt & $\delta_{X|Z}$  & \texttt{delta.XIZ} & \texttt{0}\\
	\noalign{\smallskip}
	\hline
	\multicolumn{5}{c}{\it Scatters} \\
	\noalign{\smallskip}
	\multicolumn{5}{l}{$\sigma_{Y|Z}=[\sigma_{Y|Z,0} + f_\text{knee}(Z)(\sigma_{Y|Z,0,\text{knee}}-\sigma_{Y|Z,0}) ]F_z^{\gamma_{\sigma_{Y|Z,F_z}}}D^{\gamma_{\sigma_{Y|Z,D}}}$} \\  
	 \noalign{\smallskip}  
	 Intrinsic scatter & scatter at $z=z_\text{ref}$ for $Z\ge Z_\text{knee}$	& $\sigma_{Y|Z,0}$ & \texttt{sigma.YIZ.0} & \texttt{prec.dgamma}\\
	                                            & scatter at $z=z_\text{ref}$ for $Z<Z_\text{knee}$	& $\sigma_{Y|Z,0,\text{knee}}$ & \texttt{sigma.YIZ0.knee} & \texttt{sigma.YIZ.0}\\
	 &	time evolution	with $F_z$ 			& $\gamma_{\sigma_{Y|Z,F_z}}$ & \texttt{gamma.sigma.YIZ.Fz} & \texttt{0} \\
	&	time evolution	with $D$	         	         & $\gamma_{\sigma_{Y|Z,D}}$ & \texttt{gamma.sigma.YIZ.D} & \texttt{0} \\
	\noalign{\smallskip}  
	\multicolumn{5}{l}{$\sigma_{X|Z}=\sigma_{X|Z,0} F_z^{\gamma_{\sigma_{X|Z,F_z}}}D^{\gamma_{\sigma_{X|Z,D}}}$} \\  
	 \noalign{\smallskip}  
	 Intrinsic scatter of the proxy  & scatter at $z=z_\text{ref}$	& $\sigma_{X|Z,0}$ & \texttt{sigma.XIZ.0} & \texttt{0}\\
	 & time evolution with $F_z$				& $\gamma_{\sigma_{X|Z,F_z}}$ & \texttt{gamma.sigma.XIZ.Fz} & \texttt{0}\\
	  & time evolution with $D$		& $\gamma_{\sigma_{X|Z,F_z}}$ & \texttt{gamma.sigma.XIZ.D} & \texttt{0 }\\
	\noalign{\smallskip}
	\multicolumn{5}{l}{$\rho_{XY|Z}  =  \rho_{XY|Z,0}F_z^{\gamma_{\rho_{XY|Z,F_z}}}D^{\gamma_{\rho_{XY|Z,D}}}$} \\  
	 \noalign{\smallskip}  
	 Intrinsic scatter correlation  & correlation at $z=z_\text{ref}$	& $ \rho_{XY|Z,0}$ & \texttt{rho.XYIZ.0} & \texttt{0}\\
	 & time evolution with $F_z$				& $\gamma_{\rho_{XY|Z,F_z}}$ & \texttt{gamma.rho.XYIZ.Fz} & \texttt{0}\\
	 & time evolution with $D$		& $\gamma_{\rho_{XY|Z,F_z}}$ & \texttt{gamma.rho.XYIZ.D} & \texttt{0 }\\
	 \hline  
	 \noalign{\smallskip}
	 \multicolumn{5}{c}{\it Intrinsic  distribution of the independent variable}  \\
	 \noalign{\smallskip}  
	\multicolumn{5}{l}{$p(Z) \propto \left[  \sum_k \pi_k \ {\cal N}(\mu_{Z,k} (z), \sigma_{Z,k}(z)) \right]  \text{erfc}[(\mu_{Z_\text{min}}(z)-Z)/\sigma_{Z_\text{min}}(z)/\sqrt{2}] {\cal U}(Z_\text{max}-Z)$}  \\
	 \noalign{\smallskip}  
	  Gaussian mixture & number of components   & $n_\text{mix}$ & \texttt{n.mixture} & \texttt{[1]}\\
	   & weights of the components   & $\pi_{k}$ & \texttt{pi[k]} & \texttt{ddirch }\\
	   & maximum  $Z$ value  (only for $n_\text{mix}=1$)& $Z_\text{max}$ & \texttt{Z.max} & \texttt{$+\infty$ }\\
	\multicolumn{5}{l}{$\mu_{Z,k} (z) = \mu_{Z,0k} +\gamma_{\mu_Z,F_z}T + \gamma_{\mu_Z,D}\log D$} \\ 
	 \noalign{\smallskip}  
	  Means of the Gaussian & mean of the first component at $z=z_\text{ref}$ & $\mu_{Z,01}$ & \texttt{mu.Z.0} & \texttt{dunif }\\
	  components                       & means of the additional components & $\mu_{Z,0k}$ & \texttt{mu.Z.0.mixture[k]}  & \texttt{dunif }\\
	    &  ($2\le k \le n_\text{mix}$)  at $z=z_\text{ref}$ &  &  & \\
	  & time evolution with $F_z$ & $\gamma_{\mu_Z,F_z}$ & \texttt{gamma.mu.Z.Fz} & \texttt{dt } \\
	 & time evolution with $D$    & $\gamma_{\mu_Z,D}$     & \texttt{gamma.mu.Z.D} & \texttt{dt } \\
	 \noalign{\smallskip}  
	\multicolumn{5}{l}{ $\sigma_{Z,k}(z)=\sigma_{Z,0k}F_z^{\gamma_{\sigma_{Z},F_z}}D^{\gamma_{\sigma_{Z},D}}$} \\
        \noalign{\smallskip}  
       Standard deviations of the & deviation of the first component at $z=z_\text{ref}$	& $\sigma_{Z,01}$ & \texttt{sigma.Z.0} & \texttt{prec.dgamma }\\
       Gaussian  components & deviations of the additional components 	& $\sigma_{Z,0k}$ & \texttt{sigma.Z.0.mixture[k]} & \texttt{prec.dgamma }\\
               &  ($2\le k \le n_\text{mix}$)  at $z=z_\text{ref}$ &  &  & \\
       & time evolution with $F_z$	& $\gamma_{\sigma_{Z},F_z}$ & \texttt{gamma.sigma.Z.Fz} & \texttt{0}\\
       & time evolution with $D$	& $\gamma_{\sigma_{Z},D}$ & \texttt{gamma.sigma.Z.D} & \texttt{0}\\
       \multicolumn{5}{l}{$\mu_{Z_\text{min}} (z) = \mu_{Z_\text{min,0}} +\gamma_{\mu_{Z_\text{min}},F_z}T + \gamma_{\mu_{Z_\text{min}},D}\log D$} \\ 
	 \noalign{\smallskip}  
	  Minimum $Z$ value  (only for $n_\text{mix}=1$) & typical value at $z=z_\text{ref}$ & $\mu_{Z_\text{min,0}}$ & \texttt{mu.Z.min.0} & \texttt{$-\infty$ }\\
	  & time evolution with $F_z$ & $\gamma_{\mu_{Z_\text{min}},F_z}$ & \texttt{gamma.mu.Z.min.Fz} & \texttt{dt } \\
	 & time evolution with $D$    & $\gamma_{\mu_{Z_\text{min}},D}$     & \texttt{gamma.mu.Z.min.D} & \texttt{dt } \\
	\multicolumn{5}{l}{ $\sigma_{Z_\text{min}}(z)=\sigma_{Z_\text{min,0}}F_z^{\gamma_{\sigma_{Z_\text{min}},F_z}}D^{\gamma_{\sigma_{Z_\text{min}},D}}$} \\	 \noalign{\smallskip}  
	$Z_\text{min}$  dispersion & dispersion at $z=z_\text{ref}$ & $\sigma_{Z_\text{min,0}}$ & \texttt{sigma.Z.min.0} & \texttt{0}\\
	  & time evolution with $F_z$ & $\gamma_{\sigma_{Z_\text{min}},F_z}$ & \texttt{gamma.sigma.Z.min.Fz} & \texttt{0} \\
	 & time evolution with $D$    & $\gamma_{\sigma_{Z_\text{min}},D}$     & \texttt{gamma.sigma.Z.min.D} & \texttt{0} \\
	\hline
	\end{tabular}
	}
\end{table*}

\section{Expected mass}
\label{sec_expe}

Complexity, interactions, and initial conditions can break self-similarity. We can try to model apparent deviations from determinism in terms of subtle effects. Differences from the self-similar model relating observables in X-ray and millimeter wave bands with the total mass can be efficiently expressed in a more complex picture in terms of the gas clumpiness, the gas mass fraction and the logarithmic slope of the thermal pressure profile \citep{ett15}.

Still, some inherent complexity can persist, observations may be not deep enough to assess faint features, and the model can get too intricate and unstable. One way to deal with such complexity is to introduce the concept of intrinsic scatter about a mean relation. We want to quantify the probability that a cluster has a mass given a measured proxy, e.g. the luminosity, the richness, the temperature. This can be expressed in terms of a conditional probability.

Theoretical models \citep{kai86,gio+al13,ett13}, numerical simulations \citep{sta+al10,fab+al11,ang+al12,sar+al13} and observations \citep{mau07,vik+al09} favour power-law relations between cluster properties with a lognormal scatter around the mean. We can conveniently treat the problem in terms of the normal distribution if we work in the logarithmic space. Naming the covariate (i.e. the logarithm of the proxy) and the response (i.e. the logarithm of the mass) as $X$ and $Y$, respectively, we can write, 
\beq
\label{eq_cond_1}
Y|X \sim {\cal N}(Y_X,  \sigma_{Y|X}^2),
\eeq
where ${\cal N}$ is the normal distribution, the intrinsic scatter is quantified by the standard deviation $\sigma_{Y|X}$ and the mean is linearly related to the proxy $X$,
\beq
\label{eq_cond_2}
Y_X  =  \alpha_{Y|X} + \beta_{Y|X} X + \gamma_{Y|X} T,
\eeq
where $T$($=\log F_z$) embodies the redshift dependence.

Sources of uncertainty when we try to estimate $Y$ given $X$ can come from the uncertainty associated to $X$, from uncertainties associated to the scaling relation and from uncertainties on the shape of the conditional probability.

\section{Errors and biases in model prediction}
\label{sec_delta}

Sources of uncertainty in mass prediction are variegate. In this section, we review some sources and discuss some ad hoc methods to deal with them under specific circumstances. These prescriptions cab be useful but they are not general and can fail in important cases of astronomical interest. We discuss them to illustrate the impact and the size of biases and errors which often plague the study of mass prediction and scaling relations. The general treatment of these effects in terms of a Bayesian hierarchical model, which should be used in the general case, is detailed in the following sections. In most of this section we identify the (logarithm of the) proxy with the covariate $X$ and the (logarithm of the) mass with the response $Y$.

\subsection{Intrinsic scatter}
\label{sec_delta_int_scat}

Since the scaling relation is scattered, an uncertainty related to the intrinsic dispersion $\sigma_{Y|X}$ is associated to the forecasting of the observable properties. This is usually one of the main sources of uncertainty. In the Bayesian treatment, we distinguish between a latent quantity which fits perfectly the scaling relation, $Y_X$ in Eqs.~(\ref{eq_cond_1} and \ref{eq_cond_2}), and a scattered observable quantity, $Y$ in Eq.~(\ref{eq_cond_1}), which is a scattered realization of the latent quantity $Y_X$. We associate the latent $Y_X$ with the true mass and the observable $Y$ with the WL mass.

\subsection{Statistical measurement uncertainties in the proxy}
\label{sec_delta_stat}

The statistical uncertainty $\delta_x$ in the measurement process of the proxy $X$  propagates to an uncertainty in the prediction,
\beq
\label{eq_delta_1}
\delta_Y=\beta_{Y|X} \delta_x .
\eeq
This kind of uncertainty can be reduced with better measurements but it cannot be eliminated.

\subsection{Systematic measurement uncertainties in the proxy}
\label{sec_delta_syst}

Systematic uncertainties $\Delta x$ in the proxy propagates as 
\beq
\label{eq_delta_2}
\Delta_y=\beta_{Y|X} \Delta x.
\eeq
Systematics can be related to the data-acquisition or to the statistical treatment. The former type can be eliminated only with a better understanding of the measurement process. In the following, we focus on the latter. 

A bias can be defined as the difference between the average value of the observed quantity $x$ and the true intrinsic value $X$ for objects of the same class \citep{but+al05}. If the quantity of interest is the absolute magnitude, biases are usually referred as Malmquist biases \citep{tee97}. 

These systematics can be turned on if the measurements are affected by some dispersion. This can be either the statistical uncertainty $\delta_x$ or the intrinsic scatter $\sigma_{Y|X}$. The formal treatment is the same. In the examples we detail next, we consider the statistical uncertainty. 

The two biases we review have long been known and are usually named after Eddington or Malmquist. Still these biases are not always corrected for in today analyses of scaling relations. The solutions we discuss apply only to specific cases. The general case must be addressed in a Bayesian framework, as detailed in Sec.~\ref{sec_pred}.

\subsubsection{Uneven distribution}
\label{sec_bias_edd}

\begin{figure}
       \resizebox{\hsize}{!}{\includegraphics{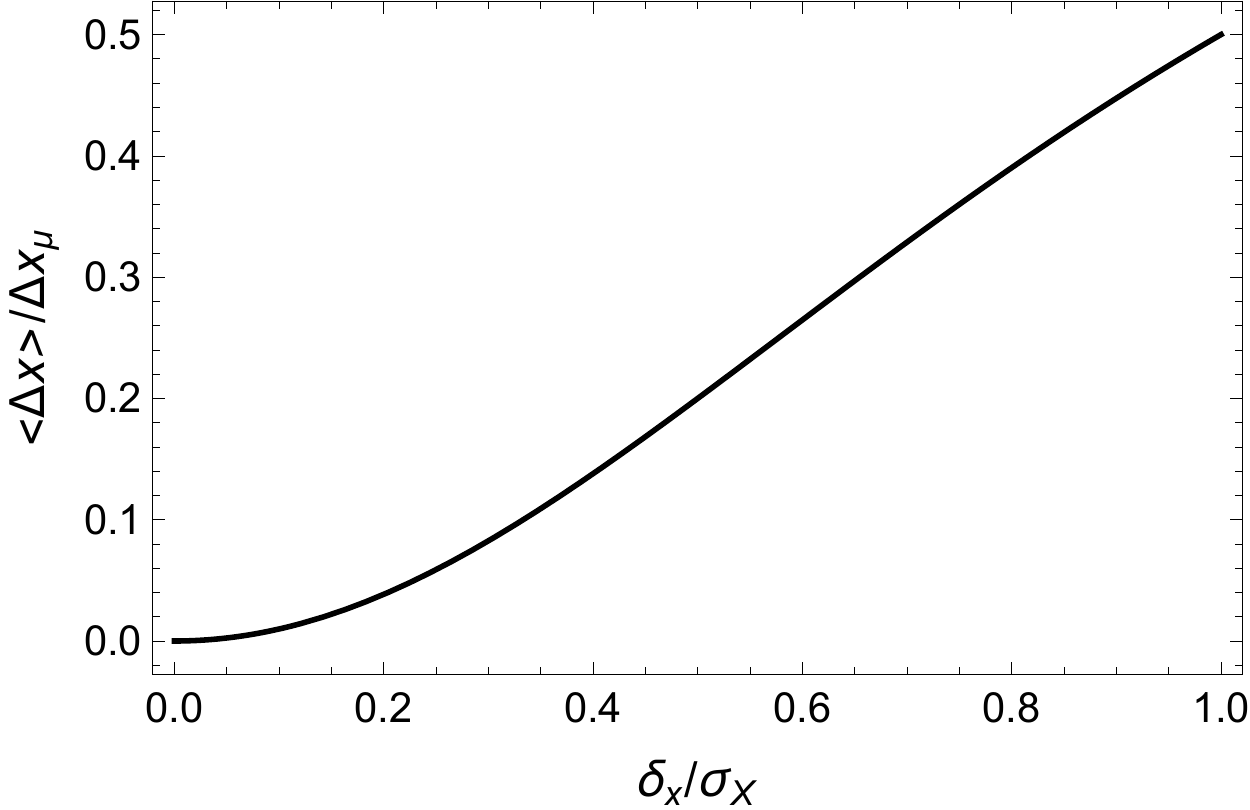}}
       \caption{The bias due to a not constant distribution of the independent variable, $p(X)$. Here, $p(X)$ is Gaussian. The bias $\langle \Delta x\rangle$, i.e. the difference between the average value of the observed quantity $x$ and the true intrinsic value $X$ for objects of the same class, is shown as a function of the measurement uncertainty $\delta_x$. The bias is in units of $\Delta x_\mu$, i.e. the distance of $x$ from the mean of the distribution; the uncertainty $\delta_x$ is in units of the standard deviation of the distribution $\sigma_X$.}
	\label{fig_bias_edd}
\end{figure}

The uneven distribution of the true variable $X$ can be a source of bias in presence of observational uncertainties $\delta_x$. Similar considerations apply in case of intrinsic scatter rather than measurements errors \citepalias{se+et15_comalit_I}.

If the true number of objects in the range $dX$ is greater than in adjacent ranges, on account of measurement errors more observations are scattered out of the range than into it \citep{edd13,edd40,jef38}. The distribution $p(x)$ of the observed variable $x$ differs from the underlaying true distribution of $X$. As a result, the average of the $X$ values of a sample of objects with fixed $x$ differs from the measured value, i.e. from $x$.

The average $X$ value of an ensemble of objects with the same measured $x$ is given by
\beq
\label{eq_bias_1}
\langle  X \rangle (x) \propto \int X \, p(x,X) d X ,
\eeq
where $p(x,X)$ is the probability of $x$ and $X$. We follow the treatment in appendix A of \citetalias{se+et15_comalit_I}. We assume that the intrinsic distribution of true values is Gaussian
\beq
\label{eq_bias_2}
X \sim {\cal N}(\mu_X, \sigma_{X}^2),
\eeq
and that the errors have a Gaussian distribution,
\beq
\label{eq_bias_3}
x|X \sim {\cal N}(X, \delta_{x}^2).
\eeq
We assume that the uncertainty $\delta_x$ is constant. A simple application of the Bayes' theorem shows that $x$ and $X|x$ are normally distributed too \citepalias{se+et15_comalit_I}. 


Given the above distributions, the mean $X$ for a given $x$ is \citepalias{se+et15_comalit_I},
\beq
\label{eq_bias_4}
\langle  X \rangle (x) =  x  + \langle \Delta x \rangle,
\eeq
with a bias
\beq
\label{eq_bias_5}
\langle \Delta x \rangle =\frac{\mu_X-x}{1+(\sigma_X/\delta_x)^2}.
\eeq
For negative (positive) slopes of the underlying distribution, i.e. $x>$$(<)\mu_X$ for a Gaussian distribution, the measured $x$ over-estimates $\langle  X \rangle (x)$, i.e. $\langle \Delta x \rangle<$$(>)0$. The bias is more pronounced at the tails, being proportional to the distance of the observed value from the mean of the distribution, $\Delta x_\mu =\mu_X-x$. The bias $\langle \Delta x \rangle$ increases with the measurement error, see Fig.~\ref{fig_bias_edd}, and decreases with the dispersion of the distribution $\sigma_X$. For $\sigma_X=\delta_x$, $\langle \Delta x \rangle = \Delta x_\mu/2$.

For a general distribution \citep{edd40},
\beq
\label{eq_bias_6}
\langle \Delta x \rangle =\delta_x^2 \frac{d \ln p(x)}{dx}.
\eeq

This kind of bias is analogous to the classical Malmquist bias \citep{mal22} and is classified as the magnitude-dependent Malmquist bias \citep{but+al05}. 

The distribution of corrected values computed according to Eq.~(\ref{eq_bias_4}) cannot be used for statistical inference without caveats \citep{edd40}. In fact, the procedure overcorrects the observational error. The distribution of the corrected values $x  + \langle \Delta x \rangle$ deviates as far from the true distribution $p(X)$ in the direction of reduced spread as the observed distribution $p(x)$ does in the direction of increased dispersion.

\subsubsection{Selection effects}
\label{sec_bias_mal}

\begin{figure}
       \resizebox{\hsize}{!}{\includegraphics{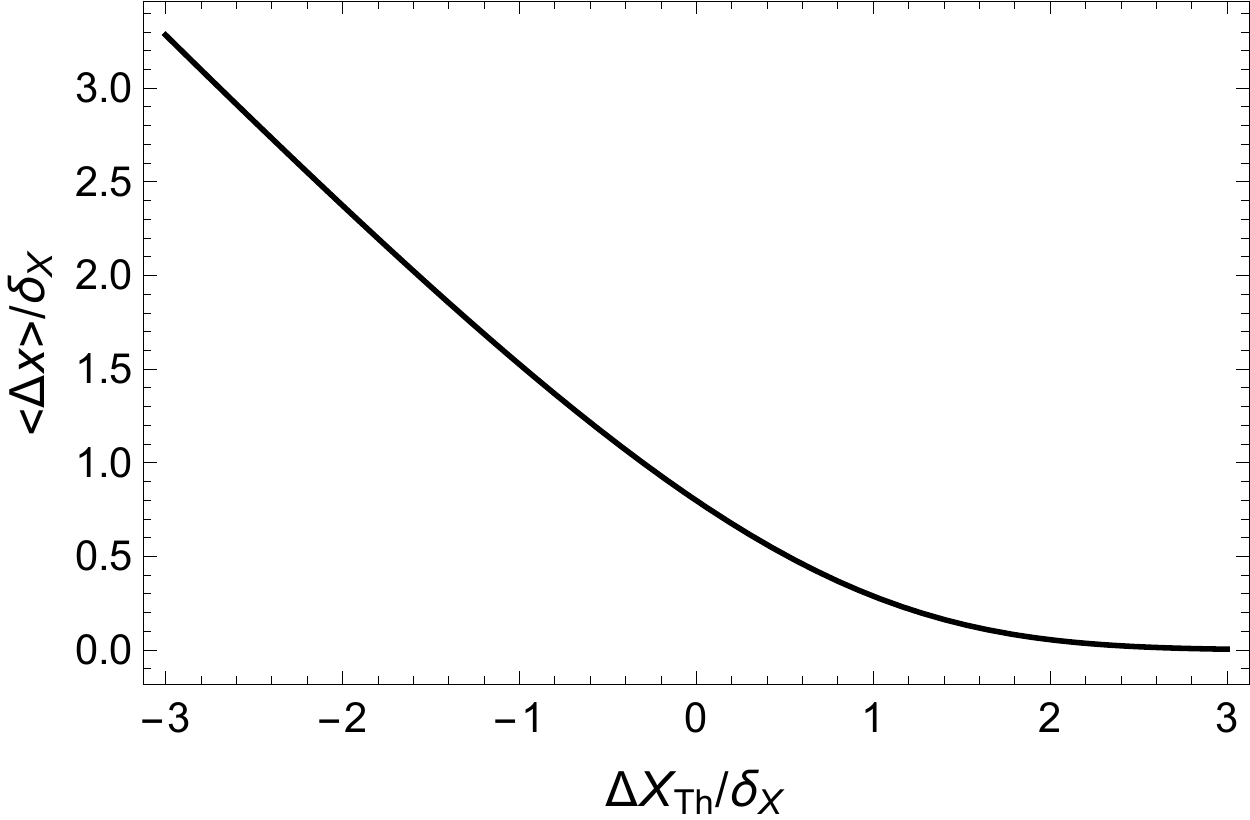}}
       \caption{The bias due to selection effects when only objects with measured $x>x_\text{th}$ are retained in the sample. The bias $\langle \Delta x\rangle$, i.e. the difference between the average value of the observed $x$ and the true intrinsic value $X_0$, is shown as a function of the distance of the true value from the threshold, $\Delta X_\text{th}=X_0-x_\text{th}$. The bias and $\Delta X_\text{th}$ are in units of the measurement uncertainty $\delta_x$.}
	\label{fig_bias_mal}
\end{figure}

A bias can show up if the conditional probability of the observed value $x$ given the true value is skewed. Selection effects can distort the distribution of observed samples. Over-luminous objects are over-represented in flux-limited samples \citep{edd14,mal20}.

Let us consider a sample selected with an hard cut in the observed value of the covariate. Only objects above a given threshold, $x>x_\text{th}$, are included. Our treatment is similar to \citet[appendix A.2]{vik+al09} and to the so called distance dependent Malmquist bias \citep{but+al05}. Due to selection effects, the conditional probability is truncated,
\beq
\label{eq_bias_7}
x|X \sim {\cal N}(X, \delta_{x}^2) {\cal U}(x-x_\text{th}),
\eeq
where $ {\cal U}$ is the step function. 

Differently from the bias discussed in Section~\ref{sec_bias_edd}, the selection bias does not come out due to the irregular distribution of the true variable. Even if the parent population of $X$ is uniformly distributed, this bias can still be in place. However, due to the bias, the distribution of the observed $X$ after selection is asymmetric. 

The average bias of the detected objects  is
\beq
\label{eq_bias_8}
\langle \Delta x \rangle = \int \int (x-X) p(x,X) dx dX.
\eeq
Let us consider a single population with a given value of $X=X_0$, wherefore the distribution can be modelled with a Dirac delta function,
\beq
\label{eq_bias_9}
X \sim \delta_\text{Dirac}(X-X_0).
\eeq
For a given $X_0$, Eq.~(\ref{eq_bias_8}) simplifies as
\beq
\label{eq_bias_10}
\langle \Delta x \rangle = \int_{x_\text{th}}^\infty (x-X_0) p(x|X_0) dx .
\eeq
We get
\beq
\label{eq_bias_11}
\langle \Delta x \rangle =\delta_x \frac{
\exp \left[   -(\Delta X_\text{th}/\delta_x)^2/2  \right]
}{
\sqrt{\pi/2}\ \text{erfc}\left[-\Delta X_\text{th}/(\sqrt{2}\delta_x)\right] 
},
\eeq
where $\Delta X_\text{th} = X_0-x_\text{th}$ and `$\text{erfc}$' is the complementary error function.

As shown in Fig.~\ref{fig_bias_mal}, the bias is prominent when the true value is deeply below the threshold, $X_0 \ll x_\text{th}$, when we expect that selected $x$'s are just above $x_\text{th}$.

Note that the correction in Eq.~(\ref{eq_bias_11}) is written in terms of the unknown $X_0$. In order to get the nominal $X_0$ given the measured $x$, the object-by-object correction should be performed by inversion,
\beq
\label{eq_bias_12}
x+\langle \Delta x \rangle(X_0)=X_0.
\eeq
However, it seems common practice to compute the correction by computing the bias for $x$ rather than for $X_0$, i.e. $X_0= x + \langle \Delta x \rangle(x)$ \citep{planck_2013_XX}. This approximation can significantly under-estimate the bias when the bias is large.

\subsection{Scaling uncertainties}
\label{sec_delta_sr}

The degree of uncertainty on the scaling relation is a source of uncertainty in the mass forecast. Scaling parameters bring their own uncertainty. Even if the proxy is accurately known, the statistical errors on the scaling parameters propagate as
\beq
\label{eq_delta_3}
\delta_Y^2 = \delta_\alpha^2 +  \delta_\beta^2 X^2+ 2 \rho_{\alpha\beta} \delta_\alpha \delta_\beta X,
\eeq
where $\rho_{\alpha\beta}$ is the uncertainty correlation factor.
 
The estimated slope and intercept are usually anti-correlated \citep{se+et15_comalit_IV}, which diminishes (augments) the combined error for positive (negative) $X$. This error associated with the scaling is proxy dependent and it can cause systematic effects.

\subsection{Scattered response}
\label{sec_delta_scat}

Most of the scaling relations on the market are not universal. We can measure the galaxy richness of a cluster in terms of the number of red-sequence galaxies in some magnitude ranges, $\lambda_\text{red}$, or in terms of galaxies with similar photometric redshifts, $\lambda_{z_\text{ph}}$. Even if we manage to obtain the true $M$--$\lambda_\text{red}$ relation, we cannot use this relation straight away to measure the mass of a cluster with known $\lambda_{z_\text{ph}}$. In fact, $\lambda_{z_\text{ph}}$ is scattered with respect to $\lambda_\text{red}$. 

This scatter $\sigma_{X_1|X_2}$, which may be unknown, can cause effects similar to measurement uncertainties in the proxy 
\beq
\Delta_Y=\beta_{Y|X_2} \sigma_{X_1|X_2},
\eeq
and the related systematic effects discussed in Sections~\ref{sec_bias_edd} and \ref{sec_bias_mal}.

\section{Model prediction}
\label{sec_pred}

The Bayesian hierarchical model developed in the CoMaLit series is already apt to model prediction. The approach is fully detailed in \citetalias{se+et15_comalit_IV} and \citet{ser16_lira}. Here, we summarize the aspects relevant to model prediction. 

Our data-set consists of a sample of clusters with measured proxies $\mathbfit{x}$. The corresponding masses $\mathbfit{y}$ may be either known, which constitute the calibration subsample, or unknown, which are the masses we want to predict.

\subsection{General case}
\label{sec_pred_gene}

In the general CoMaLit scheme, we distinguish between the result of the measurement $y_i$ of the $i$-th cluster, the true value of the observable $Y_i$, i.e. the measurement we would get in absence of noise and which can differ from the actual measurement for the observational error $\delta_{y,i}$, and the $Y_{Z,i}$ variable, which exactly fits the scaling relation and can differ from $Y_i$ for the intrinsic scatter. Similar notations hold for $x_i$, $X_i$, and $X_{Z,i}$. $Z$ is the underlying latent variable which $X_{Z}$ and $Y_{Z}$ are linearly related to.

The full probability density can be written as
\begin{multline}
\label{eq_com_1}
p(\mathbfit{x},\mathbfit{y},\mathbfit{X},\mathbfit{Y},\mathbfit{Z},{\mbox{\boldmath $\pi$}}_\text{SR},{\mbox{\boldmath $\pi$}}_\text{Z}) =
\prod_i^{N_\text{obs}} {\cal L}_i (\mathbfit{X}_i,\mathbfit{Y}_i,{\mbox{\boldmath $\pi$}}_\text{SR}) P(\mathbfit{Z}_i,{\mbox{\boldmath $\pi$}}_\text{Z}) \\
\times 
P({\mbox{\boldmath $\pi$}}_\text{SR}) P({\mbox{\boldmath $\pi$}}_\text{Z}),
\end{multline}
where ${\mbox{\boldmath $\pi$}}_\text{SR}$ and ${\mbox{\boldmath $\pi$}}_\text{Z}$ are the meta-parameters characterizing the scaling relation and the intrinsic distribution of the independent variable, respectively. $P({\mbox{\boldmath $\pi$}}_\text{SR})$ and $P({\mbox{\boldmath $\pi$}}_\text{Z})$ are the prior distributions. $N_\text{obs}$ is the number of clusters in the sample and the independent variable $\mathbfit{Z}_i$ is drawn from the probability distribution of the selected clusters, $P(Z, {\mbox{\boldmath $\pi$}}_\text{Z})$. 

The likelihood for the $i$-th cluster can be expressed as
\beq
\label{eq_com_2}
{\cal L}_i=P(\mathbfit{x}_i,\mathbfit{y}_i | \mathbfit{X}_i,\mathbfit{Y}_i)P(\mathbfit{X}_i,\mathbfit{Y}_i | \mathbfit{Z}_i, {\mbox{\boldmath $\pi$}}_\text{SR}),
\eeq
where $P(\mathbfit{x}_i,\mathbfit{y}_i | \mathbfit{X}_i,\mathbfit{Y}_i)$ is the joint conditional probability of the measurements given the true values and $P(\mathbfit{X}_i,\mathbfit{Y}_i | \mathbfit{Z}_i, {\mbox{\boldmath $\pi$}}_\text{SR})$ is the probability of the true values given the latent variable $Z$. 

The expressions of the full probability density in Eq.~(\ref{eq_com_1}) and the likelihood in Eq.~(\ref{eq_com_2}) are general. The form of the conditional probabilities $P(\mathbfit{x}_i,\mathbfit{y}_i | \mathbfit{X}_i,\mathbfit{Y}_i)$ and $P(\mathbfit{X}_i,\mathbfit{Y}_i | \mathbfit{Z}_i, {\mbox{\boldmath $\pi$}}_\text{SR})$ can change in presence of selection effects which can truncate or curtail the distributions. As a result, the sampled population of the covariate, which models the distribution of the selected objects, differs from the parent population of all objects before selection.

In the CoMaLit approach, we model these probabilities as bivariate normal distributions to accounts for correlated measurements errors and correlated scatters. Distributions can be truncated to account for selection effects. We refer to \citetalias{se+et15_comalit_IV} for full details.

For linear scaling and in absence of biases, the expected value of $Y$ given $Z$  is
\beq
\label{eq_com_3}
Y_Z  =  \alpha_{Y|Z}+\beta_{Y|Z} Z + \gamma_{Y|Z} T ,
\eeq
where $T$ expresses the redshift-dependence and it is assumed to be known. A similar relation holds between $X_Z$ and $Z$.

\subsection{Unscattered proxy}
\label{sec_pred_simpl}

We can then distinguish different cases in astronomy. In forecasting, we may want to estimate $Y$ given $X$, which we can directly measure. For scaling relations, we may be interested either in the relation between $Y$ and $Z$, which is the latent variable which ideally fits a scaling relation, or in the relation between $Y$ and $X$. In the next sections, we will focus on forecasting, i.e. $Y$ given $X$.

When we are interested in forecasting, we can usually deal with a simplified scheme. If we want to predict the mass corresponding to the actual realization of the proxy, we can consider the proxy as unscattered. The proxy is still affected by measurement uncertainties but, in this simplified forecasting scheme, we do not have to consider the conditional scatter of the independent variable. In the CoMaLit formalism, we can now identify $X$ with $Z$, i.e. $X$ is not affected by intrinsic scatter with respect to $Z$. 

For mass prediction, $y_i$ is the (logarithm of the) measured mass, e.g., the WL mass, $Y_i$, is the true WL mass and the $Y_{Z,i}$ is the true mass. We usually want to estimate $Y_{Z,i}$, i.e. we are mostly interested in the true mass rather than in the WL mass. The measured mass proxy is $x_i$ whereas the proxy we would measure in absence of noise is $X_i$. The full probability density in Eq.~(\ref{eq_com_1}) reduces to 
\begin{multline}
\label{eq_com_4a}
p(\mathbfit{x},\mathbfit{y},\mathbfit{X},\mathbfit{Y},{\mbox{\boldmath $\pi$}}_\text{SR},{\mbox{\boldmath $\pi$}}_\text{X}) =
\prod_i^{N_\text{obs}} {\cal L}_i (\mathbfit{X}_i,\mathbfit{Y}_i,{\mbox{\boldmath $\pi$}}_\text{SR}) P(\mathbfit{X}_i,{\mbox{\boldmath $\pi$}}_\text{X}) \\
\times 
P({\mbox{\boldmath $\pi$}}_\text{SR}) P({\mbox{\boldmath $\pi$}}_\text{X}),
\end{multline}
where the likelihood can be written as 
\beq
\label{eq_com_4}
{\cal L}_i=P(\mathbfit{x}_i,\mathbfit{y}_i | \mathbfit{X}_i,\mathbfit{Y}_i)P(\mathbfit{Y}_i | \mathbfit{X}_i, {\mbox{\boldmath $\pi$}}_\text{SR}) .
\eeq
The intrinsic distribution of the covariate can be written as $P(\mathbfit{X}_i | {\mbox{\boldmath $\pi$}}_\text{X})$ rather than in terms of $Z$, which is now an unnecessary variable.

Here, we are limiting our analysis only to the $N_\text{obs}$ selected clusters which are included in the sample. Because of the catalog selection method, the sample only includes $N_\text{obs}$ clusters out of the possible $N_\text{tot}$ sources located within the survey area. $N_\text{tot}$ is unknown and should thus also be treated as a parameter of the statistical model. The data likelihood should be written as a product running over all the $N_\text{tot}$ clusters, either detected or missed. Here, $N_\text{tot}$ is expressed in terms of the theoretical halo mass function whereas the selection function enters as the probability of including a source with a given $\mathbfit{x}$ and $\mathbfit{y}$.

However, when the sample selection function is dependent of the observed $\mathbfit{x}$ only, i.e. it does not depend on $\mathbfit{y}$, and the measurement errors on $\mathbfit{y}$ and $\mathbfit{x}$ are statistically independent, the terms in the data-likelihood including $\mathbfit{y}$ and $\mathbfit{Y}$ of the $N_\text{tot}-N_\text{obs}$ missing sources can be integrated over \citep{kel07}. This is actually the case we deal with in the next sections. We refer to \citet{kel07} for full details.

As a result, the final data-likelihood can be written in term of the $N_\text{obs}$ detected clusters only, with the caveat that $p(\mathbfit{X})$ only models the distribution of the clusters that have been included in the sample. This simplification is expressed in Eq.~(\ref{eq_com_4}). Because we are primarily interested in inference on regression parameters and unknown masses of the detected clusters, we can omit from the likelihood the lasting terms depending on $N_\text{tot}$.

This approach is workable only if we are not interested in cosmology and cluster abundance, since $N_\text{tot}$ embodies the cosmological information.

If the mass $y_i$ is measured and known, the variables $Y_i$ and $Y_{Z,i}$ are constrained in two ways. On one side, the response is anchored to the measurement result through $P(\mathbfit{x}_i,\mathbfit{y}_i | \mathbfit{X}_i,\mathbfit{Y}_i)$, where we are considering possibly correlated measurements. On the other side, $Y_i$ has to fit the scaling relation through $P(\mathbfit{Y}_i | \mathbfit{X}_{i}, \pi_\text{SR})$. This double requirement allows us to estimate at the same time $X$, $Y$, $Y_Z$ and the scaling relation. Whereas $y$ is the direct output of the measurement process, $Y$ is a refined estimate of the WL mass which exploits the information about linearity.

If the data is missing, $Y$ is determined solely by the scaling relation. The measured $x_i$ and the missing $y_i$ are uncorrelated,
\beq
\label{eq_com_5}
P(\mathbfit{x}_i,\mathbfit{y}_i | \mathbfit{X}_i,\mathbfit{Y}_i)= P(\mathbfit{x}_i | \mathbfit{X}_i) P(\mathbfit{y}_i | \mathbfit{Y}_i) . 
\eeq
The true WL mass $\mathbfit{Y}_i$ and the true mass $\mathbfit{Y}_{Z,i}$ are now determined by the scaling $P(\mathbfit{Y}_i | \mathbfit{X}_i, {\mbox{\boldmath $\pi$}}_\text{SR})$. The constraint from $x_i$ still actively participates in the inference of the unbiased intrinsic distribution $P(X)$. 

The scaling parameters are mainly determined by the data with measured $y$. However, the full distribution $p(x)$ helps to constrain the covariate distribution $p(X)$, which is crucial for bias correction.

In the CoMaLit approach, we can model the distribution of the covariate $P(Z)$ as either a Gaussian distribution, a mixture of Gaussian distributions, or a truncated Gaussian distribution. The truncation can be hard or smooth, when the Gaussian is multiplied by a complementary error function, see \citetalias{se+et15_comalit_IV} and \citet{ser16_lira} for details. All the parameters entering the covariate distribution can be redshift-evolving. This set of distributions can properly model an ample range of physical cases. The most basic case, i.e. a simple Gaussian function, is flexible enough to cover most of the sensible cases.

The described Bayesian hierarchical model can be useful for mass estimates even if the data-set is complete with no missing measurement. In fact, the regression determines both the WL calibrated mass $Y$ and the true mass $Y_Z$. This can be particularly useful in presence of selection biases. On the other hand, if the WL measurement is missing, the regression can estimate the mass based on the global scaling.

Scaling parameters and missing data are determined at the same time during the regression. The Bayesian method accounts for heteroscedastic and possibly correlated measurement errors, intrinsic scatter, time evolution, distribution of the independent variable and selection effects in the covariate or the response variable. In the same way, the procedure provides corrected estimates of the proxy $X$ too.

The Bayesian treatment deals with the scaling relation and the forecasting at the same time and the predicted masses do not suffer from the issues raised in Section~\ref{sec_delta}. Systematics are accounted for whereas all statistical sources of uncertainty are included in the uncertainty budget.

The CoMaLit approach fully implements the described scheme by modelling the conditional probabilities with redshift evolving normal distributions. Gaussian distributions can be truncated or skewed to account for selection effects. The priors are fully customizable.

The method is implemented in \texttt{LIRA}, and it does not formally distinguish between data and parameters. Missing data values are treated as parameters to be estimated. If a node, i.e. a statistical relation $P(\mathbfit{y}_i | \mathbfit{Y}_i)$, contains a missing value for the unobserved element ($\mathbfit{y}_i=\texttt{NA}$), the posterior distribution for the node can be sampled. The resulting samples at the \texttt{NA} locations in the response vector are samples from the posterior predictive distribution and can be summarized just like model parameters.

\begin{table}
\caption{Characteristics of the considered WL samples. Col.~1: name. Col.~2: number of clusters, $N_\mathrm{cl}$ . Cols.~3 and 4: typical redshift and dispersion. Cols.~5 and 6: typical mass and dispersion. Col.~7: main reference. Typical values and dispersions are computed as bi-weighted estimators. Masses are in units of $10^{14}M_\odot$.}
\label{tab_samples}
\resizebox{\hsize}{!} {
\begin{tabular}{ l r r r r r l}     
Sample		 &	$N_\mathrm{Cl}$	&	$z$	&	$\sigma_z$	&	$M_{500}$	&	$\sigma_M$ &  reference \\ 
\hline
LC$^2$-single	&	506	        &	0.31	&	0.18	&	5.0	&	4.5	&	\citet{ser15_comalit_III}	 \\
CCCP  		        &	54    		&	0.24	&	0.10	&	7.7	&	4.2	&	\citet{hoe+al15}          \\
WtG  		        &	51    		&	0.38	&	0.14	&	10.5	&	5.1	&	\citet{wtg_III_14}	\\
LoCuSS  		        &	50    		&	0.23	&	0.04	&	6.5	&	3.1	&	\citet{ok+sm16}	          \\
CLASH		        &	20		&	0.38	&	0.13	&	10.0	&	4.0	&	\citet{ume+al16b} 	  \\
\hline	
\end{tabular}
}
\end{table}

\section{Calibration sample}
\label{sec_samp}

Scaling relations to be used in mass forecasting have to be assessed through a subsample of clusters with well measured masses. Weak lensing masses are arguably the more reliable mass estimates \citepalias{se+et15_comalit_I}. They do not depend on the dynamical state of the cluster (as far as the mass distribution is well modelled), are nearly unbiased, and with a quite small intrinsic scatter of $\sim 10$ per cent. These features make WL masses optimal for the calibration of mass proxies \citep{be+kr11,ras+al12,se+et15_comalit_I}.

As calibration sample, we considered the Literature Catalogs of weak Lensing Clusters of galaxies (LC$^2$), the largest compilation of WL masses up to date\footnote{The catalogues are available at \url{http://pico.bo.astro.it/\textasciitilde sereno/CoMaLit/LC2/}.}  \citepalias{ser15_comalit_III}. LC$^2$ are standardized catalogues of galaxy clusters with measured WL mass retrieved from literature. The latest compilation lists 879 weak lensing analyses of clusters and groups with measured redshift and mass from 81 bibliographic sources. The catalogs report coordinates, redshift, WL masses to over-densities of 2500, 500, 200, and to the virial radius and spherical WL masses within 0.5, 1.0, and 1.5 Mpc.

For our main analysis, we considered the catalogue of unique entries LC$^2$-single. For clusters with multiple analyses, the WL study exploiting the most deep observations and the most complete multi-band coverage was picked out. This criterion usually selects the most recent analysis, e.g. either the CLASH \citep[Cluster Lensing And Supernova survey with Hubble,][]{ume+al16b} or the WtG \citep[Weighing the Giants,][]{wtg_III_14} clusters. 

The catalogue spans a large interval in redshift $0.02\ls z\ls 1.5$ and mass $0.03\ls M_{500}/(10^{14} M_\odot)\ls 30$. Main properties are summarized in Table~\ref{tab_samples}. We refer to \citetalias{ser15_comalit_III} for a detailed discussion of the catalog properties. For testing and comparison, we also considered smaller but homogeneous sample, see Table~\ref{tab_samples}.

\section{The {\it Planck} SZ catalog}
\label{sec_planck}

The second {\it Planck} Catalogue of Sunyaev-Zel'dovich Sources \citep[PSZ2,][]{planck_2015_XXVII} is the largest SZ-selected sample of galaxy clusters yet produced and the deepest all-sky catalogue of galaxy clusters. The second release of the catalogue contains 1653 sources detected with a SNR (signal-to-noise ratio) above 4.5 from the 29 months full-mission data, of which 1203 are confirmed clusters with counterparts identified in external optical or X-ray samples or by dedicated follow-ups.

As selection pipeline, we considered the Matched Multi-Filter method MMF3, which identified 1271 candidates. The redshift determination is available for 926 clusters. The catalog spans a broad nominal mass range from $M_{500}\sim0.1$ to $16\times10^{14}M_\odot$. The median redshift is $z\sim 0.22$. We provide WL calibrated mass estimates for the subsample of MMF3 detections with measured redshift.

\subsection{{\it Planck} mass prediction}
\label{sec_planck_SR}

Let us summarize the mass calibration adopted by the {\it Planck} team \citep{planck_2013_XX}. The {\it Planck} team estimated the mass of the detected clusters with known redshift assuming the best-fitting scaling relation between $M_{500}$ and $Y_{500}$, i.e. the spherically integrated Compton parameter within a sphere of radius $r_{500}$. The scaling between mass and SZ flux was modelled as \citep{planck_2013_XX},
\beq
\label{eq_scal_rela}
E_z^{-2/3} \left[ \frac{D_z^2Y_{500}}{10^{-4}\mathrm{Mpc}^2}\right]=10^{\alpha}\left[ \frac{M_{500}}{M_\mathrm{pivot}}\right]^\beta,
\eeq
where the redshift evolution is set to self-similarity, $D_z$ is the angular diameter distance to the cluster, $M_\mathrm{pivot}=6\times10^{14}M_\odot$, $\alpha=-0.186 \pm0.011$ and $\beta=1.79\pm0.06$.

The $Y_{500}$--$M_{500}$ relation was determined through multiple steps \citep{planck_2013_XX}. The first step involved $Y_\text{X}$, i.e. the X-ray analogue of $Y_{500}$ defined as the product of the gas mass within $r_{500}$ and the spectroscopic temperature outside the core \citep{kra+al06}\footnote{The X-ray quantity $Y_\text{X}$ defined in this section should not be confused with the variable $Y_X$ used in the Bayesian model.}. The local $Y_\text{X}$--$M_{\text{HE},500}$ relation was estimated in a sample of 20 relatively relaxed local clusters \citep{arn+al10}. This calibration sample is not representative of the full {\it Planck} catalogue. Masses estimated through this scaling relation are denoted $M_{500}^{Y_\text{X}}$. 

Secondly, the $Y_{500}$--$M_{500}^{Y_\text{X}}$ relation was computed for 71 detections from the {\it Planck} cosmological sample with good quality XMM-Newton observations. The SZ fluxes were re-estimated within a sphere of radius $r^{Y_\text{X}}_{500}$, centred on the position of the X-ray peak. 

The $Y_{500}$--$M_{500}$ relation, see Eq.~(\ref{eq_scal_rela}), was finally used as a prior that cuts through the parameter likelihood contours of the SZ observables to estimate the mass. This mass should equal the hydrostatic mass expected for a cluster consistent with the assumed scaling relation, at a given redshift and given the {\it Planck} posterior information. In the {\it Planck} papers, the masses estimated through the scaling relation are denoted as $M_\text{SZ}$ or $M_{500}^{Y_z}$.

The errors quoted in the PSZ2 catalog are the 68 per cent confidence statistical errors and are based on the measurement uncertainties only. They do not include the statistical errors on the scaling relation, the intrinsic scatter in the relation, or systematic errors in data selection for the scaling relation fit. 

A bias $b_\text{SZ}$ can still persist in the HE mass measurements, $M_{\text{HE},500}=(1-b_\text{SZ})M_{500}$ \citep{planck_2013_XX}. Based on a suite of numerical simulations \citep{bat+al12,kay+al12}, the {\it Planck} team estimated $b_\text{SZ}=0.2^{+0.1}_{-0.2}$.

\begin{table}
\caption{Bias of the {\it Planck} SZ masses derived from various calibration samples of WL masses. Col.~1: sample name. Col.~2: number of WL clusters, $N_\text{cl}$, detected by {\it Planck} with the MMF3 algorithm. Cols.~3 and 4: typical redshift and dispersion. Cols.~5 and 6: typical WL mass and dispersion in units of $10^{14}M_\odot$. Col.~7: bias $b_\text{SZ}=\ln(M_\text{SZ}/M_\text{WL})$. Typical values and dispersions are computed as bi-weighted estimators.}
\label{tab_bias_MSZ}
\resizebox{\hsize}{!} {
\begin{tabular}{ l r c c r c r@{$\,\pm\,$}l}     
Sample		 &	$N_\text{Cl}$	&	$z$	&	$\sigma_z$	&	$M_{500}$	&	$\sigma_{M_{500}}$  &\multicolumn{2}{c}{$b_\text{SZ}$} \\ 
\hline
LC$^2$-single	        &	135	        &	0.24	&	0.14	&	7.8	&	4.8	& $-$0.25 &	0.04 \\
CCCP  		        &	35    		&	0.23	&	0.07	&	8.5	&	3.8	& $-$0.22	&	0.07	\\
CLASH		        &	13		&	0.37	&	0.13	&	11.3	&	3.3	& $-$0.39	&	0.08	\\
LoCuSS  		        &	38    		&	0.23	&	0.04	&	7.5	&	2.8	& $-$0.18 &	0.05	\\
WtG  		        &	37    		&	0.36	&	0.13	&	11.5	&	5.2	& $-$0.43 &	0.06	\\
\hline	
\end{tabular}
}
\end{table}

\subsection{{\it Planck} masses vs WL masses}
\label{sec_planck_bias}

The level of bias of the {\it Planck}-derived masses can be assessed by direct comparison with WL masses. We identified a WL cluster as a counterpart of a PSZ2 candidate if their redshifts differ for less than $\Delta z =0.01$ and their angular separation in the sky does not exceed 3/2 times the SZ position uncertainty\footnote{The positional accuracy quoted in the MMF3 algorithm catalogue is defined as the the 95 per cent confidence interval of the distribution of radial displacement \citep{planck_2015_XXVII}.}. We also excluded cluster pairs separated by more than 10 arcminutes. In case of multiple matches, we retained the candidate which was the closest in the sky.

The estimate of the bias is strongly dependent on the methodology and on the calibration sample.  According to the sample selection criteria, corrections for Eddington bias have to be applied (\citetalias{se+et15_comalit_I}, \citealt{bat+al16}). The usual approach, which we followed to easy comparison with previous results, is to consider only the subsample of SZ detected clusters with WL observations. 

Alternatively, you can consider {\it Planck} measurements as follow-up observations of the weak-lensing sample and then account for non-detections by setting the SZ signal of non-detected clusters to values corresponding to a multiple of the average SZ error \citep{bat+al16}. 

For a detailed discussion of the bias and of recent measurements, we refer to \citetalias{ser+al15_comalit_II}. Here, we only update the results to the latest catalogue releases. To assess the effect of the calibration sample, we measured the bias for {\it Planck} clusters with estimated WL masses. Following the approach detailed in \citetalias{se+et15_comalit_I}, the bias was estimated by fitting the relation\footnote{The bias $b_\text{SZ}$ in Eq.~(\ref{eq_bias_13}) is defined as $b_\text{SZ}=\ln M_\text{SZ} -\ln M_\text{WL}$ whereas in the {\it Planck} papers the bias is defined as $b_\text{SZ}=M_\text{SZ}/M_{500}-1$. For low values of $b_\text{SZ}$ and being the WL mass a low scatter proxy of the mass, the two definitions are approximately interchangeable.}
\beq
\label{eq_bias_13}
\ln \langle M_\text{SZ} \rangle = b_\text{SZ} + \ln \langle M_\text{WL} \rangle.
\eeq
We considered both observational uncertainties and intrinsic scatters in the two mass estimates. The regression was performed with \texttt{LIRA}. We found $b_\text{SZ}=-0.25\pm0.14$, see Table~\ref{tab_bias_MSZ}. 

Our result is consistent with previous estimates based on WL comparison. \citet{lin+al14} found a large bias of $b_\text{SZ}=0.30\pm0.06$ in the WtG sample. \citet{planck_2015_XXIV} found a bias of $b_\text{SZ}=-0.32\pm0.07$ for the WtG sample, $b_\text{SZ}=-0.22\pm0.09$ for the CCCP \citep[Canadian Cluster Comparison Project,][]{hoe+al15} sample and $b_\text{SZ}\sim 1$ from CMB (Cosmic Microwave Background) lensing. \citet{smi+al16} found that the mean bias with respect to the LoCuSS \citep[Local Cluster Substructure Survey,][]{ok+sm16} sample is $b_\text{SZ}=-0.05\pm0.04$.

\subsection{WL calibrated masses}
\label{sec_planck_WL}

\begin{table*}
\caption{The first 10 entries of the catalogue of WL calibrated masses of the PSZ2 clusters, \textsc{HFI\_PCCS\_SZ-MMF3\_R2 .08\_MWLc.dat}. The full catalog is available in electronic form. Col.~1: index of detection. Col.~2: name of PSZ2 detection. Cols.~3 and 4: Right ascension and declination (J2000) of the {\it Planck} detection. Col. 5: redshift of the cluster; Cols.~6 and 7: WL calibrated mass $M_\text{WLc}$ and associated error. Cols.~7 and 8: WL mass of the LC$^2$-single counterpart, $M_\text{LC2}$, and associated error. Cols.~9 and 10: SZ mass proxy $M_\text{SZ}$ and associated error, as from the {\it Planck} catalog. Masses are within $r_{500}$ and they are in units of $10^{14}M_\odot$.}
\label{tab_PSZ2_MWLc}
\resizebox{\hsize}{!} {
\begin{tabular}{llrrrrrrrrr}     
\texttt{INDEX}  &   \texttt{NAME}  & \texttt{RA}  & \texttt{DEC} & \texttt{z}  & \texttt{MWLc}   &  \texttt{MWLc\_ERR} &  \texttt{MLC2}  &   \texttt{MLC2\_ERR} &  \texttt{MSZ}   &   \texttt{MSZ\_ERR}\\
\hline
1  & PSZ2\_G000.04+45.13 & 229.1905 &-1.0172  &0.1198 &4.175  &0.512   &{\texttt NA}     &{\texttt NA}     &3.776  &0.361 \\
2  & PSZ2\_G000.13+78.04 & 203.5587 &20.2560  &0.1710 &5.897  &0.478   &{\texttt NA}     &{\texttt NA}     &5.121  &0.335 \\
3  & PSZ2\_G000.40-41.86 & 316.0845 &-41.3542 &0.1651 &5.593  &0.473   &9.387  &2.485  &4.785  &0.338 \\
4  & PSZ2\_G000.77-35.69 & 307.9728 &-40.5987 &0.3416 &8.192  &0.883   &{\texttt NA}     &{\texttt NA}     &6.310  &0.599 \\
6  & PSZ2\_G002.08-68.28 & 349.6324 &-36.3326 &0.1400 &2.681  &0.737   &{\texttt NA}     &{\texttt NA}     &2.789  &0.417 \\
8  & PSZ2\_G002.77-56.16 & 334.6595 &-38.8794 &0.1411 &5.006  &0.437   &5.731  &1.673  &4.389  &0.300 \\
9  & PSZ2\_G002.82+39.23 & 235.0152 &-3.2851  &0.1533 &6.221  &0.616   &9.100  &1.950  &5.490  &0.495 \\
11 & PSZ2\_G003.21-76.04 & 358.3512 &-33.2932 &-1.    &{\texttt NA}     &{\texttt NA}      &{\texttt NA}     &{\texttt NA}     &{\texttt NA}     &{\texttt NA}    \\
12 & PSZ2\_G003.91-42.03 & 316.4675 &-38.7532 &0.1521 &5.151  &0.482   &{\texttt NA}     &{\texttt NA}     &4.525  &0.350 \\
13 & PSZ2\_G003.93-59.41 & 338.6081 &-37.7413 &0.1510 &7.471  &0.496   &5.467  &2.004  &6.652  &0.257 \\
\hline	
\end{tabular}
}
\end{table*}

\begin{figure}
       \resizebox{\hsize}{!}{\includegraphics{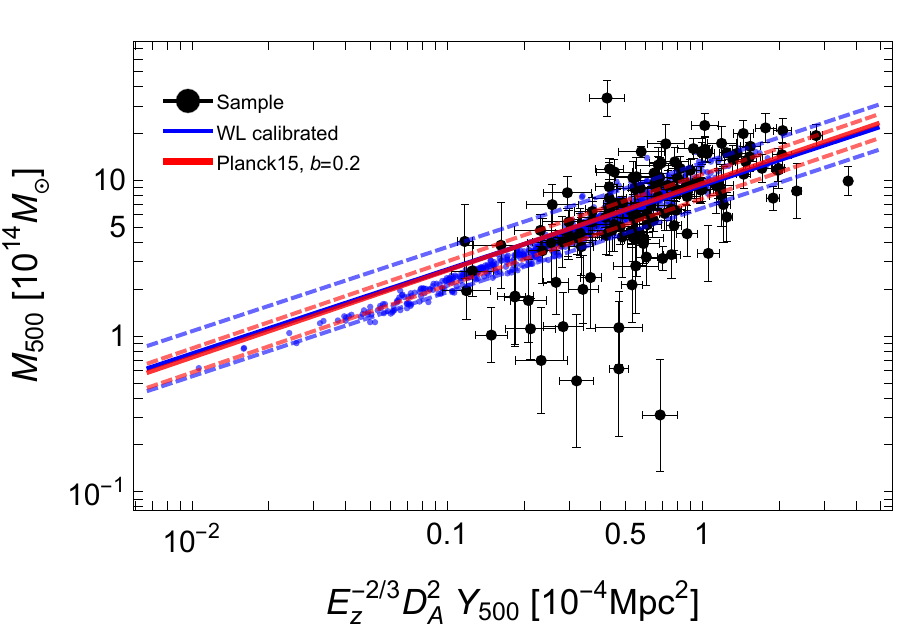}}
       \caption{SZ flux versus mass. The black points with error bars picture the calibration sample, $M_\text{WL}$. The blue points picture the clusters with predicted masses $M_\text{WLc}$ at the position of the measured SZ flux. The dashed blue lines show the median scaling relation (full blue line) plus or minus the intrinsic scatter  at $z=0.24$. The full red line shows the {\it Planck} derived relation assuming a bias $b_\text{SZ}=-0.2$. The dashed red lines show the {\it Planck} relation for null bias (lower bound) and for $b_\text{SZ}=-0.4$ (upper bound).}
	\label{fig_YSZ_M500}
\end{figure}

\begin{figure}
       \resizebox{\hsize}{!}{\includegraphics{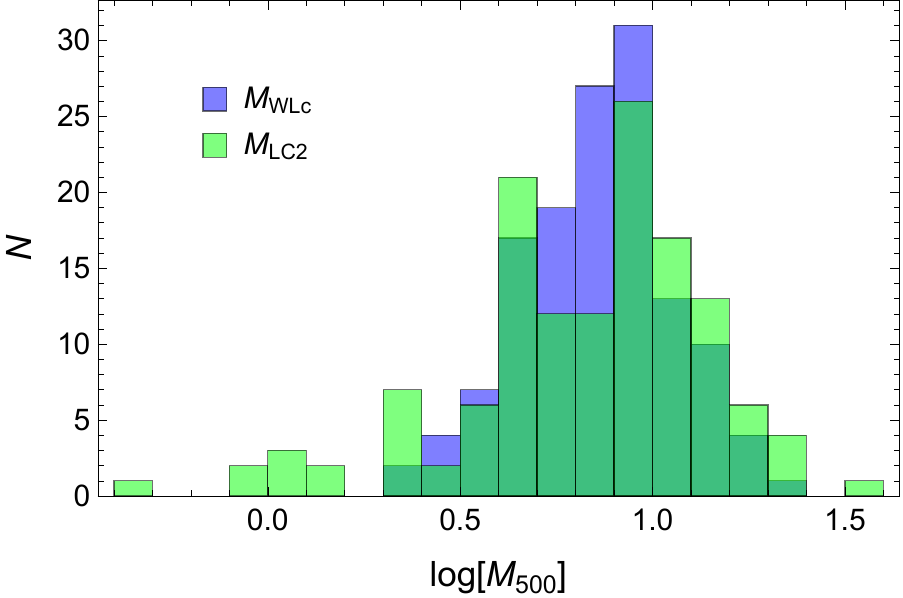}}
       \caption{Histograms of the logarithm of the WL calibrated mass ($M_\text{WLc}$, blue charts) and of the original masses $M_\text{LC2}$ (green charts) for the calibration sample in the analysis of the {\it Planck} clusters. Masses are in units of $10^{14}M_\odot$.}
	\label{fig_MLC2_MWLc_histo}
\end{figure}

\begin{figure}
       \resizebox{\hsize}{!}{\includegraphics{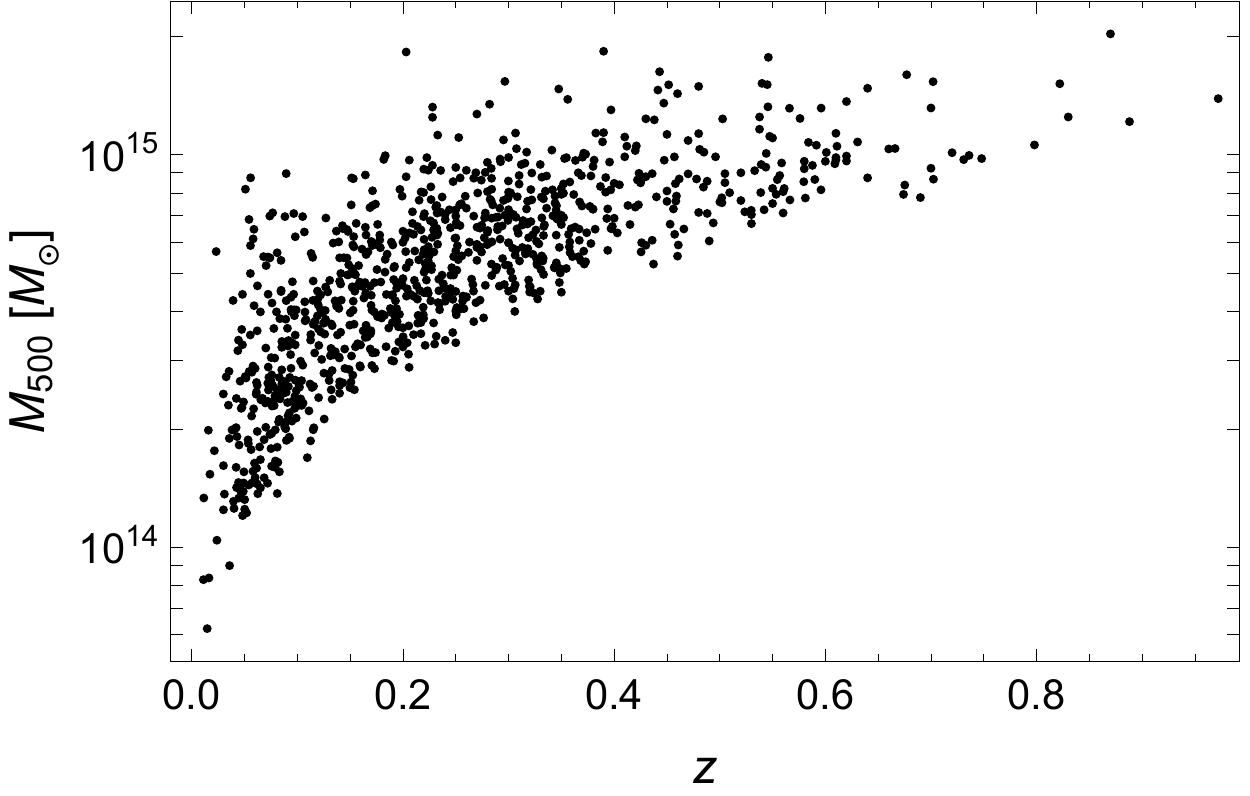}}
       \caption{Distribution of the 926 PSZ2-MMF3 clusters with known redshift in the $M_{500}$-$z$ plane. Masses are WL calibrated.}
	\label{fig_z_MWLc_planck}
\end{figure}

Following the method described in Section~\ref{sec_pred_simpl}, we derived the WL calibrated masses for all the MMF3 {\it Planck} clusters with measured redshift. The calibration sample consists of 135 clusters with WL measurements, see Table~\ref{tab_bias_MSZ}. The catalogue of masses is released with the paper and it is also available at \url{http://pico.bo.astro.it/\textasciitilde sereno/CoMaLit}. An extract of the catalog is presented in Table~\ref{tab_PSZ2_MWLc}. Here and in the following we denote direct WL mass measurements, i.e. the original masses measured with WL analyses and not reprocessed through the Bayesian method, as $M_\text{WL}$. Masses from the LC$^2$-single catalog can be alternatively denoted as $M_\text{WL}$ or $M_\text{LC2}$. We denote weak lensing calibrated (WLc) masses, i.e. mass estimates based on the scaling relation and on the WL measurement, if any, as $M_\text{WLc}$. 


\subsubsection{The proxy}

Scaling relations referring to an over-density radius can be ambiguous. If we know the over-density radius, which we need to measure the proxy, we already know the mass by definition. Here we favour a more direct approach wherein the proxy is measured in an effective over-density radius $r_{\Delta,\text{proxy}}$ whereas the mass is measured within the true over-density radius $r_\Delta$. The effective $r_{\Delta,\text{proxy}}$ can be estimated with a scaling relation or can be a fixed length. These choices simplify the proxy, whose measurement is then not correlated to the mass.

As a proxy, we considered the spherically integrated Compton parameter $Y_{500}$\footnote{The values of $Y_{500}$ were retrieved from the third extension HDU (Header Data Unit) of the individual MMF3 algorithm catalogue. Practically, we translated the quoted masses to $Y_{500}$ by means of the scaling relation. We tested that this estimate is fully compatible with the SZ flux derived by us from the likelihood tabulated function applying the scaling-relation prior.}. The effective $r_{500,\text{proxy}}$ was determined by the {\it Planck} team through the $Y_{500}$-$M_{500}$ relation, see Sec. \ref{sec_planck_SR}.

The Compton parameter $Y_{500}$ probes a large region and it is only marginally dependent on the integration radius. We tested that its estimate varies negligibly if we re-derived it from the likelihood under a WL prior, i.e. adopting $r_{500}$ from the WL analysis, or after re-centring \citepalias{ser+al15_comalit_II}.

In the CoMaLit notation, we can identify $x$ with the measured proxy,
\beq
\label{eq_planck_MWL_1}
x=\log(10^4 D_{z}^2 Y_{500}/ E_{z}^{2/3}/\text{Mpc}^2).
\eeq

\subsubsection{Selection effects}

The PSZ2 sample is selected by the SZ properties. By approximating $\text{SNR}\sim Y_{500}/\delta Y_{500}$, the limiting flux of the $i$-th {\it Planck} cluster is obtained by multiplying the minimum SNR by the uncertainty on the integrated Compton parameter \citepalias{ser+al15_comalit_II}. In terms of SNR,
\beq
\label{eq_planck_MWL_4}
x_{\mathrm{th},i}=x_i +\log ( \text{SNR}_\text{th}/ \text{SNR}_i ),
\eeq
where the limiting signal-to-noise ratio is $\text{SNR}_\text{th}=4.5$. The selection threshold affects the conditional probability of the measurements, which we model as in Eq.~(\ref{eq_bias_7}),
\beq
x_i|X_i \sim {\cal N}(X_i, \delta_{x,i}^2) {\cal U}(x_i-x_{\mathrm{th},i}).
\eeq
We refer to \citetalias{se+et15_comalit_IV} and \citet{ser16_lira} for full details. 

Since the selection thresholds are point-dependent, there is no steep lower bound in the distribution of the observed SZ fluxes. Furthermore, the distribution is smoothed by observational errors. As a result, the distribution of the covariate is regular and can be well approximated as a redshift-evolving Gaussian function. We refer to \citetalias{ser+al15_comalit_II} and \citetalias{se+et15_comalit_IV} for details. 

We considered the full PSZ2 and we did not apply selection procedures based on the WL properties. Masses of clusters without WL estimates were treated as missing data. Clusters with known WL mass effectively sample the massive end of the PSZ2 catalog \citepalias{ser+al15_comalit_II}. Under the assumption that there are no breaks at the lower mass end or at high redshifts, the calibration sample follows the same scaling relation of the full sample and it is affected by the same intrinsic scatter. As far as we model the distribution of the proxies of the full sample, selection effects related to the WL measurements can be neglected if the scatter and the scaling of the WL clusters are representative of the full PSZ2 sample.

\subsubsection{The masses}

The response is the mass. We denote
\begin{align}
y     &= \log(M_{\text{LC2},500}/10^{14}/M_\odot)\ , \label{eq_planck_MWL_2} \\
Y_X     &= \log(M_{\text{WLc},500}/10^{14}/M_\odot)\ . \label{eq_planck_MWL_2b}
\end{align}

We linearly fitted the proxies and the responses. The fitting procedure retrieves at the same time the masses and the scaling relation.

The scattered scaling relation can be expressed as
\beq
\label{eq_planck_MWL_5}
Y \sim {\cal N}(Y_X= \alpha_{Y|X} +\beta_{Y|X}X +\gamma_{Y|X}T,\ \sigma_{Y|X}^2).
\eeq
The parameter $\gamma_{Y|X}$ accounts for the redshift-evolution of the relation. Since the proxies are rescaled by $E_z^{2/3}$, see Eq.~(\ref{eq_planck_MWL_1}), $\gamma_{Y|X}=0$ means self-similar evolution.

Since the intrinsic scatter of the scaling is expected to be log-normal, we fitted the logarithms. When needed, here and in the following, we assumed relations derived for the log-normal distribution to go from $\langle \log y \rangle$ to $\log \langle y \rangle$ (and similarly for other variables).

The results of the regression are plotted in Fig.~\ref{fig_YSZ_M500}. As scaling parameters, we found $\alpha_{Y|X}=0.89\pm0.04$,  $\beta_{Y|X}=0.54\pm0.06$, $\gamma_{Y|X}=1.57\pm0.48$ and $\sigma_{Y|X}=0.14\pm0.02$. The above scaling could be used to predict masses only with the caveats discussed in Sec.~\ref{sec_delta}.

We stress that Eq.~(\ref{eq_planck_MWL_5}) is functional to mass estimates. It cannot be compared to self-similar models or results from numerical simulations since it does not deal with the intrinsic scatter in the SZ flux \citepalias{ser+al15_comalit_II}. We do not need to model this scatter if we are only interested in mass prediction.

For a detailed discussion of the $Y_{500}$--$M_{500}$ scaling we refer to \citetalias{ser+al15_comalit_II} and  \citetalias{se+et15_comalit_IV}. 

Due to the regularization process and the bias correction, the distribution of WL calibrated masses is more regular than that of the original WL masses from the calibration sample, see Fig.~\ref{fig_MLC2_MWLc_histo}. Measurements at either the very high or the very low end of the mass spectrum are viewed as unlikely and the tails of the $M_\text{WLc}$ distribution are less pronounced.

The distribution of {\it Planck} selected clusters with known counterparts with redshift in the mass-redshift ($M_{500}$-$z$) plane, using the WL calibrated proxy, is shown in Fig.~\ref{fig_z_MWLc_planck}. Apart from the shift towards higher masses, the distribution agrees with the results in \citet{planck_2015_XXVII}.

\subsection{WL vs. SZ calibration}
\label{sec_planck_WL_vs_SZ}

\begin{figure}
       \resizebox{\hsize}{!}{\includegraphics{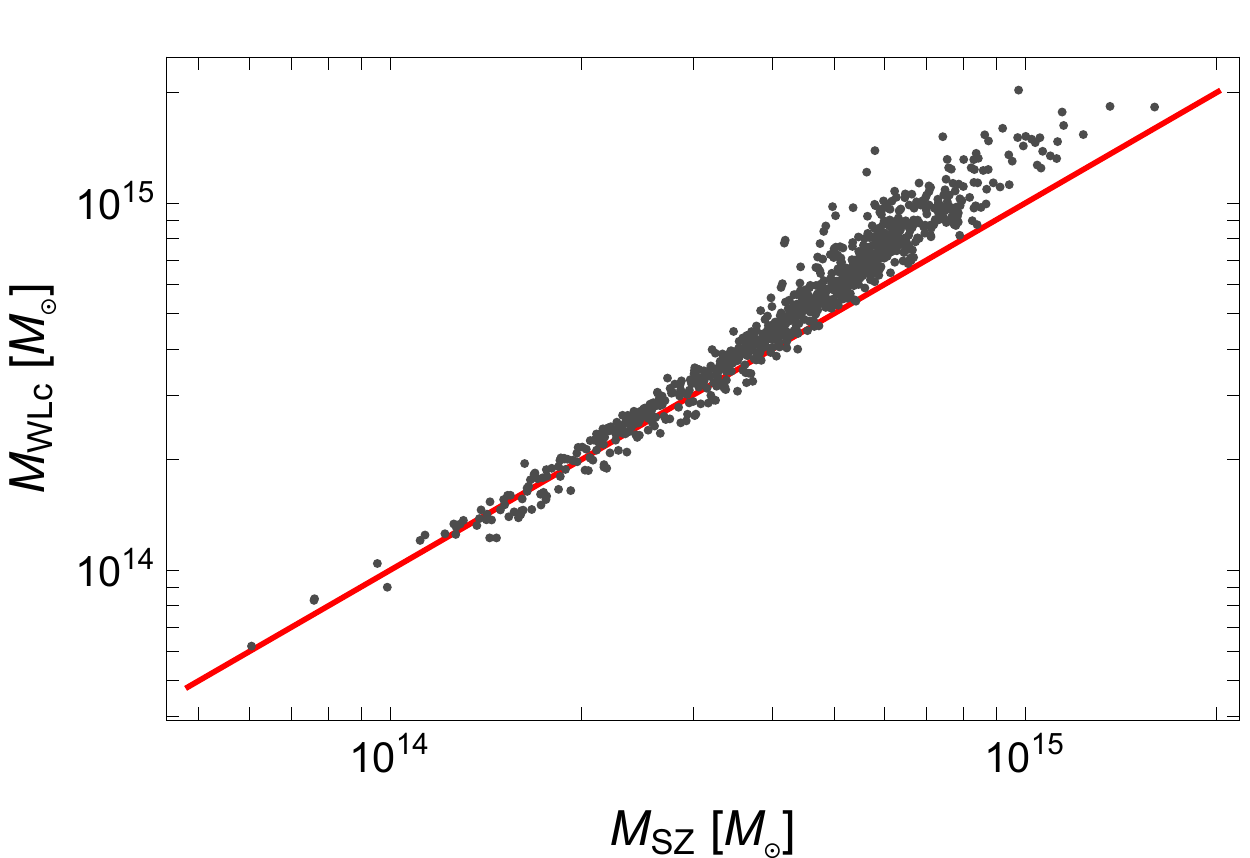}}
       \caption{{\it Planck} SZ masses $M_\text{SZ}$ versus WL calibrated masses $M_\text{WLc}$ for the MMF3 clusters with measured redshift. Masses are in units of $M_\odot$. The red line shows the bisection $M_\text{SZ}=M_\text{WLc}$.}
	\label{fig_MSZ_MWLc}
\end{figure}

\begin{figure}
       \resizebox{\hsize}{!}{\includegraphics{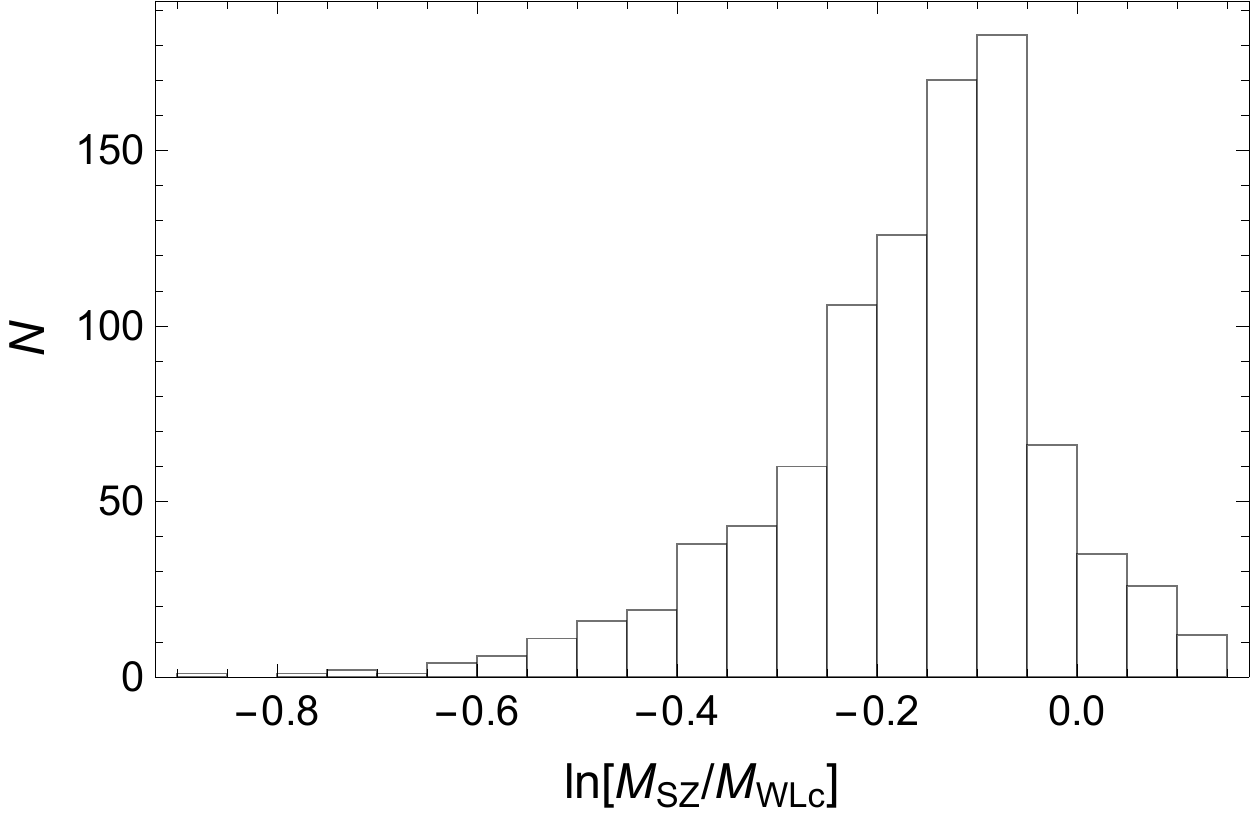}}
       \caption{Histogram of the logarithm of the ratio of the {\it Planck} SZ masses $M_\text{SZ}$ to the WL calibrated masses $M_\text{WLc}$ for the MMF3 clusters with measured redshift.}
	\label{fig_MSZ_MWLc_histo}
\end{figure}

The computed WL calibrated masses differ substantially from the SZ masses, see Figs.~\ref{fig_MSZ_MWLc} and \ref{fig_MSZ_MWLc_histo}. We can look for mass or redshift dependences of the bias in the full sample. 

Analogously to Sec.~\ref{sec_planck_bias}, the bias for the full {\it Planck} sample can be computed by comparison with the WL calibrated masses, $M_\text{WLc}$. The full sample includes {\it Planck} clusters either with or without WL counterparts. We found $b_\text{SZ}=-0.17\pm0.01$, in line with the bias computed for the subsample of {\it Planck} clusters in the LC$^2$-single by comparison with the direct $M_\text{WL}$ measurements, see Table~\ref{tab_bias_MSZ}. 

The {\it Planck} clusters with WL counterparts only probe the high-mass tail of the sample. The agreement between the two estimates suggests that the bias is not mass dependent. This conclusion seems to be at odds with Fig.~\ref{fig_MSZ_MWLc}, where the discrepancy between WL and SZ calibrated masses is more pronounced for massive clusters. However, this feature is driven by the high-mass clusters preferentially selected at high redshift. We then considered both the mass and the redshift dependence,
\beq
\ln \langle M_\text{SZ} \rangle =b_\text{SZ} + \beta_\text{SZ} \ln \langle M_\text{WLc} \rangle +\gamma_\text{SZ} T ,
\eeq
where $T=\log F_z$. We found $b_\text{SZ}=0.05 \pm0.11$, $\beta_\text{SZ}=0.94\pm0.08$ and $\gamma_\text{SZ}=-2.40\pm0.63$. This confirms that the bias is nearly mass independent but it is significantly more pronounced at high redshifts.

\section{The SDSS redMaPPer catalogue}
\label{sec_redmapper}

\begin{table*}
\caption{The first 10 entries of the catalogue of WL calibrated masses of the redMaPPer clusters, \textsc{redmapper\_dr8\_public\_v6.3\_MWLc.dat}. The full catalog is available in electronic form. Col.~1: index of detection. Col.~2: name of redMaPPer detection. Cols.~3, 4 and 5: right ascension, declination (J2000) and redshift. Cols.~6 and 7: WL calibrated mass, $M_\text{WLc}$, and associated error. Cols.~8 and 9: WL mass of the LC$^2$-single counterpart, $M_\text{LC2}$, and associated error. Masses are within $r_{200}$ and they are in units of $10^{14}M_\odot$.}
\label{tab_redmapper_MWLc}
\begin{tabular}{llrrrrrrr}        
\texttt{INDEX}  &   \texttt{NAME}          &  \texttt{RA}  &   \texttt{DEC}  &   \texttt{z}    &  \texttt{MWLc}    &  \texttt{MWLc\_ERR} &  \texttt{MLC2}  &   \texttt{MLC2\_ERR} \\
\hline
1    &   RMJ155820.0+271400.3 & 239.580  &27.233 &  0.0948 & 15.324 & 1.732  &   8.777  & 1.476   \\
2    &   RMJ164019.8+464241.5 & 250.080  &46.712 &  0.2328 & 17.670 & 1.829  &   11.729 & 1.852   \\
3    &   RMJ131129.5-012028.0 & 197.870  &-1.341 &  0.1824 & 14.793 & 1.428  &   15.033 & 1.025   \\
5    &   RMJ090912.2+105824.9 & 137.300  &10.974 &  0.1705 & 15.635 & 1.619  &   7.842  & 2.537   \\
6    &   RMJ133520.1+410004.1 & 203.830  &41.001 &  0.2317 & 16.801 & 1.692  &   14.957 & 1.864   \\
7    &   RMJ003208.2+180625.3 & 8.034    &18.107 &  0.3976 & 21.265 & 2.578  &   {\texttt NA}     & {\texttt NA}      \\
8    &   RMJ224319.8-093530.9 & 340.830  &-9.592 &  0.4410 & 25.295 & 3.468  &   20.294 & 3.865   \\
9    &   RMJ100214.1+203216.6 & 150.560  &20.538 &  0.3237 & 13.457 & 1.115  &   {\texttt NA}     & {\texttt NA}      \\
11   &   RMJ082529.1+470800.9 & 126.370  &47.134 &  0.1277 & 11.968 & 1.164  &   {\texttt NA}     & {\texttt NA}      \\
13   &   RMJ091753.4+514337.5 & 139.470  &51.727 &  0.2269 & 13.785 & 1.226  &   10.371 & 1.121   \\

\hline	
\end{tabular}
\end{table*}

\begin{figure}
       \resizebox{\hsize}{!}{\includegraphics{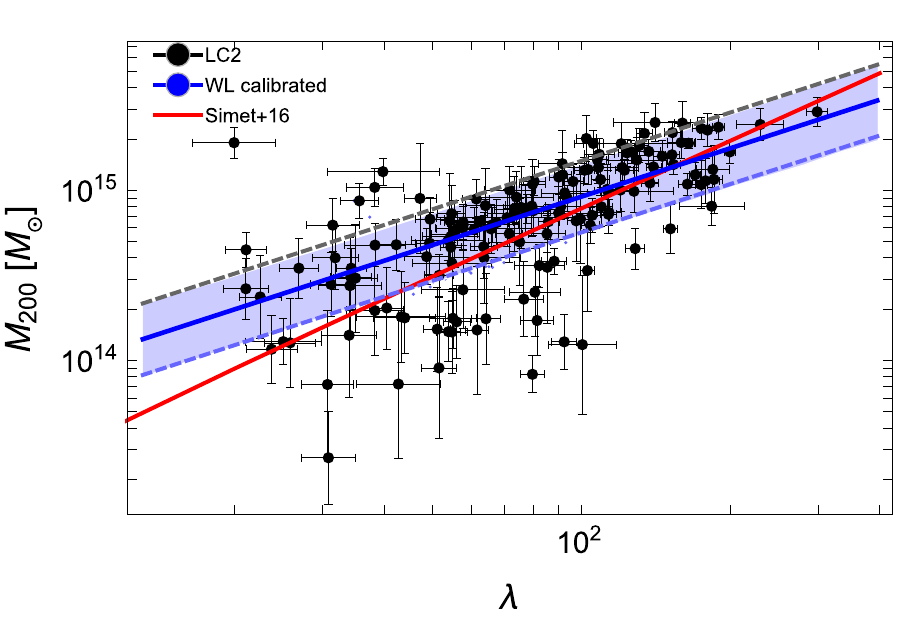}}
       \caption{Optical richness $\lambda$ vs mass. The black points with error bars picture the calibration sample, $M_\text{LC2}$. The blue points picture the clusters with predicted masses $M_\text{WLc}$ at the position of the measured $\lambda$. The dashed blue lines show the median scaling relation (full blue line) plus or minus the intrinsic scatter at $z=0.37$. The shaded blue region encloses the $68$ per cent confidence region around the median relation due to uncertainties on the scaling parameters. The red line is the relation adapted from \citet{sim+al17a}.}
	\label{fig_lambda_M200}
\end{figure}

\begin{figure}
       \resizebox{\hsize}{!}{\includegraphics{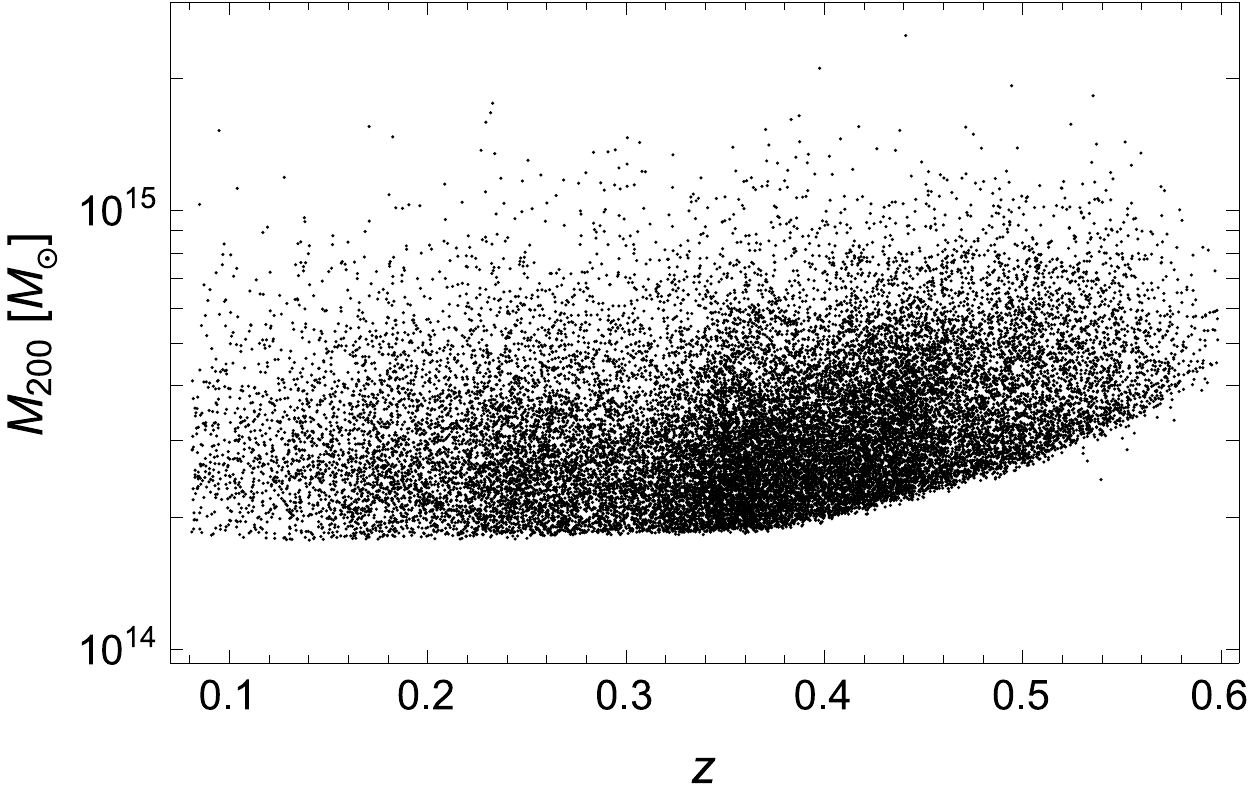}}
       \caption{Distribution of the redMaPPer clusters in the $M_{200}$-$z$ plane. Masses are WL calibrated.}
	\label{fig_z_MWLc_redmapper_v63}
\end{figure}

Optical richness or luminosity of galaxy clusters as measured from large surveys are often used as a mass proxy. \citet{wen+al15} collected and corrected cluster mass measurements based on X-ray or SZ observations and calibrated the rescaled total $r$-band luminosity to derive the mass of the clusters in the WHL \citep{wen+al12} catalog. 

Here, we consider the SDSS redMaPPer (red-sequence Matched-filter Probabilistic Percolation) catalogue, which was compiled by applying a red-sequence cluster finder to the DR8 release of the SDSS (Sloan Digital Sky Survey), covering $\sim10000~\deg^2$ \citep{ryk+al14}. The resulting catalog\footnote{The latest version of the catalog at the time of writing is v6.3. It is available at \url{http://risa.stanford.edu/redMaPPer/}.} contains $\sim 26000$ candidate clusters over the redshift range $0.08 \ls z \ls 0.55$. When available, we considered the spectroscopic redshift. Otherwise, we used the photometric redshift.

The WL mass is available for 149 redMaPPer clusters. We identified counterparts in the LC$^2$-single catalogue by matching pairs whose redshifts differ by less than $\Delta z =0.05$ and whose angular separation in the sky does not exceed 5 arcminutes.

\subsection{The proxy}

The redMaPPer optical richness $\lambda$ of a cluster is defined as the sum of the probabilities of the galaxies found near a cluster to be actually cluster members. All galaxies above a cut-off luminosity and below a distance radial cut which scales with richness are considered. According to our notation, the proxy is $x=\log\lambda$

\subsection{Selection effects}

Clusters are included in the catalog if $\lambda$ exceeds 20 times the scale factor $S_\mathrm{RM}$, which is a function of the photometric redshift of the cluster. This criterion approximately selects clusters with at least 20 galaxy counts above the flux limit of the survey or the cut-off luminosity at the cluster redshift, whichever is higher \citep{ryk+al14}. We denote the threshold for the $i$-th proxy as
\beq
\label{eq_redmapper_MWL_1}
x_{\mathrm{th},i}=\log(20 S_{\mathrm{RM},i}).
\eeq 

The effect of the selection threshold is twofold. Firstly, it skews the conditional probability of the observed value, $x$, given the true value, $X$, see Sec.~\ref{sec_bias_mal}. This is modelled as 
\beq
p(x|X) \propto {\cal N}(X, \delta_{x}^2) {\cal U}(x-x_\text{th}).
\eeq

Secondly, since the threshold is approximately constant (for most of the clusters $S_{\mathrm{RM},i} \sim 1$), the intrinsic distribution of richnesses is skewed too. We model the distribution of the proxies as a smoothly truncated Gaussian,
\beq
P(X) \propto {\cal N}\left(\mu_X,\ \sigma_{X}^2 \right)  \mathrm{erfc} \left( \frac{\mu_\mathrm{\chi}-X}{\sqrt{2}\sigma_\mathrm{\chi}}\right),
\eeq
where the low value tail of the normal distribution ${\cal N}$ is suppressed by the complementary error function\footnote{In the notation of Table~\ref{tab_par},
$\mu_\mathrm{\chi}=\mu_{Z_\text{min}}$ and $\sigma_\mathrm{\chi}=\sigma_{Z_\text{min}}$.}. In our regression analysis, we treated the effective threshold $\mu_\mathrm{\chi}$ as known and time independent. We fixed it at the minimum of the measured $x_{\mathrm{th},i}$. 

Alternatively, we also considered a threshold which is constant and equal to the minimum of the measured $x_{\mathrm{th},i}$ up to $z\sim 0.4$ and then steeply increases with $z$ to mimic the time dependence of the lower envelope of the observed richnesses\footnote{This feature is not available in the public release of the {\tt LIRA} package.}. We checked that variations in final results are negligible and that this parametrization does not improve the modelling of the distribution of richness as a function of redshift. In the following, we will discuss the reference case only.

The transition length $\sigma_\mathrm{\chi}$ was left as a free, time-independent parameter of the regression.  

Mean $\mu_X$ and standard deviation $\sigma_{X}$ of the normal distribution can evolve with redshift as formalized in \citetalias{se+et15_comalit_IV}.

\subsection{The masses}

We considered WL masses within $r_{200}$, which better matches the region analysed by the redMaPPer algorithm. The catalog of WL calibrated masses is released with the paper. An extract is presented in Table~\ref{tab_redmapper_MWLc}. The redMaPPer catalogue of WL calibrated masses spans one decade in mass from $M_{200}\sim 2 \times 10^{14} M_\odot$ to  $\sim 2.5 \times 10^{15} M_\odot$. The typical mass is $M_{200}\sim 3.1 \times 10^{14} M_\odot$ with a dispersion of $\sigma_M \sim 10^{14} M_\odot$. It samples the high mass end of the halo mass function up to redshift $z=0.55$.

The fitted conditional probability for the calibration sample is
\beq
\label{eq_redmapper_MWL_2}
Y \sim {\cal N}(Y_X= \alpha_{Y|X} +\beta_{Y|X}X +\gamma_{Y|X}T,\ \sigma_{Y|X}^2),
\eeq
with $\alpha_{Y|X}=-0.92\pm0.17$,  $\beta_{Y|X}=0.94\pm0.09$, $\gamma_{Y|X}=-0.16\pm0.55$ and $\sigma_{Y|X}=0.21\pm0.02$.

WL calibrated masses are shown in Fig.~\ref{fig_lambda_M200}. Due to the cut-off in richness, most of the redMaPPer clusters are beyond the peak of the richness distribution. As discussed in Sec.~\ref{sec_bias_edd}, these measured $\lambda$ are likely over-estimated. In the plot of the predicted $M_\text{WLc}$ versus the observed richness in Fig.~\ref{fig_lambda_M200}, they lie on the right of the estimated scaling relation, which aligns the true values of mass and richness.

The distribution of redMaPPer selected clusters in the mass-redshift ($M_{200}$-$z$) plane, using the WL calibrated proxy, is shown in Fig.~\ref{fig_z_MWLc_redmapper_v63}. The mass limit of the sample is nearly constant at $M_{200} \gs 2.1 \times 10^{14}M_\odot$ up to $z\sim0.4$ and then slowly increases up to the redshift limit of the sample. This flatness mirrors the selection properties of the redMaPPer catalog, which is nearly complete at $z\ls 0.3$ for $\lambda \gs 30$ \citep{ryk+al14}.

\subsection{Note added}

Mass calibration of redMaPPer clusters has been very recently addressed by a couple of papers, one of them posted to the public archive just before \citep{far+al16} and the other one just after \citep{sim+al17a} the first submission of this paper.

\citet{far+al16} calibrated the mass of redMaPPer clusters with a stacked spectroscopic analysis of simulated galaxy surveys. They found $\beta_{Y|X} = 1.31 \pm 0.06_\text{stat}\pm 0.13_\text{sys}$ at redshift $z=0.2$, slightly steeper than our result.

\citet{sim+al17a} performed a measurement of the mass-richness relation of the redMaPPer galaxy cluster catalogue using stacked weak lensing data of of 5570 cluster in the redshift range $0.1\la z\la0.33$ from the Sloan Digital Sky Survey. They found a power-law index of $1.33\pm0.09$.

This result is marginally consistent with ours, see Fig.~\ref{fig_lambda_M200}. The relations nearly agree at the very massive end. However, our derived scaling relation, see Eq.~(\ref{eq_redmapper_MWL_2}), is shallower and it corresponds to more massive clusters at the low richness end.

The source of disagreement may be the different treatment of the Eddington bias, illustrated in Sec.~\ref{sec_bias_edd}, which is very significant for the redMaPPer clusters. In fact, the peak of the richness distribution is just beyond the threshold $\lambda \sim 20$. Due to error measurements most of the measured richnesses are then larger than the true richnesses. The measured richnesses are consistently on the right side with respect to the scaling, see Fig.~\ref{fig_lambda_M200}. The bias is more significant at the smaller richnesses, where the relative errors are larger. To account for this, \texttt{LIRA} can model the richness distribution as a redshift-evolving normal probability density smoothly truncated at the low end.

This means that the different scaling relations cannot be compared straight away. Our scaling relation refers to the true richness, whereas the scaling relation in \citet{sim+al17a} refers to the measured richness. The comparison to be fair, we should not compare the scaling relations but the forecasted masses. In a sense, the scaling in \citet{sim+al17a} can be suitable for forecasting, but it is not representative of the `true' scaling relation. On the contrary, to use our scaling relation for forecasting we should plug in the unbiased richnesses, see Sec.~\ref{sec_delta}.

\section{MCXC}
\label{sec_mcxc}

\begin{table*}
\caption{The first 10 entries of the catalogue of WL calibrated masses of the MCXC clusters, \textsc{mcxc\_MWLc.dat}. The full catalog is available in electronic form. Col.~1: index of detection. Col.~2: name of the detection. Cols.~3 and 4: right ascension and declination (J2000) of the MCXC cluster. Col. 5: redshift of the cluster; Cols.~6 and 7: WL calibrated mass $M_\text{WLc}$ and associated error. Cols.~7 and 8: WL mass of the LC$^2$-single counterpart, $M_\text{LC2}$, and associated error. Col.~9: X-ray mass proxy $M_\text{X}$ from the MCXC catalog. Masses are within $r_{500}$ and they are in units of $10^{14}M_\odot$.}
\label{tab_mcxc_MWLc}
\resizebox{\hsize}{!} {
\begin{tabular}{llrrrrrrrrr}     
\texttt{INDEX}  &   \texttt{NAME}  & \texttt{RA}  & \texttt{DEC} & \texttt{z}  & \texttt{MWLc}   &  \texttt{MWLc\_ERR} &  \texttt{MLC2}  &   \texttt{MLC2\_ERR} &  \texttt{MX}   &   \texttt{MX\_ERR}\\
\hline
1  & MCXC\_J0000.1+0816 & 0.030 & 8.274   & 0.0396 & 0.757 & 0.130 &  {\texttt NA}    & {\texttt NA}     & 0.737  & {\texttt NA} \\  
2  & MCXC\_J0000.4-0237 & 0.103 & -2.625  & 0.0379 & 0.342 & 0.077 &  {\texttt NA}    & {\texttt NA}     & 0.330   &{\texttt NA} \\  
3  & MCXC\_J0001.6-1540 & 0.412 & -15.681 & 0.1246 & 1.792 & 0.211 &  {\texttt NA}    & {\texttt NA}     & 1.656   &{\texttt NA} \\  
4  & MCXC\_J0001.9+1204 & 0.488 & 12.073  & 0.2033 & 3.073 & 0.277 &  {\texttt NA}    & {\texttt NA}     & 2.693   &{\texttt NA} \\  
5  & MCXC\_J0003.1-0605 & 0.799 & -6.086  & 0.2320 & 6.062 & 0.471 &  5.886 & 1.214  & 5.219   &{\texttt NA} \\  
6  & MCXC\_J0003.2-3555 & 0.801 & -35.927 & 0.0490 & 1.236 & 0.175 &  {\texttt NA}    & {\texttt NA}     & 1.202   &{\texttt NA} \\  
7  & MCXC\_J0003.8+0203 & 0.961 & 2.063   & 0.0924 & 1.834 & 0.217 &  {\texttt NA}    & {\texttt NA}     & 1.734   &{\texttt NA} \\  
8  & MCXC\_J0004.9+1142 & 1.247 & 11.701  & 0.0761 & 1.364 & 0.184 &  {\texttt NA}    & {\texttt NA}     & 1.301   &{\texttt NA} \\  
9  & MCXC\_J0005.3+1612 & 1.344 & 16.211  & 0.1164 & 2.674 & 0.270 &  {\texttt NA}    & {\texttt NA}     & 2.493   &{\texttt NA} \\  
10 & MCXC\_J0006.0-3443 & 1.513 & -34.724 & 0.1147 & 2.904 & 0.284 &  {\texttt NA}    & {\texttt NA}     & 2.712   &{\texttt NA} \\
\hline	
\end{tabular}
}
\end{table*}

\begin{figure}
       \resizebox{\hsize}{!}{\includegraphics{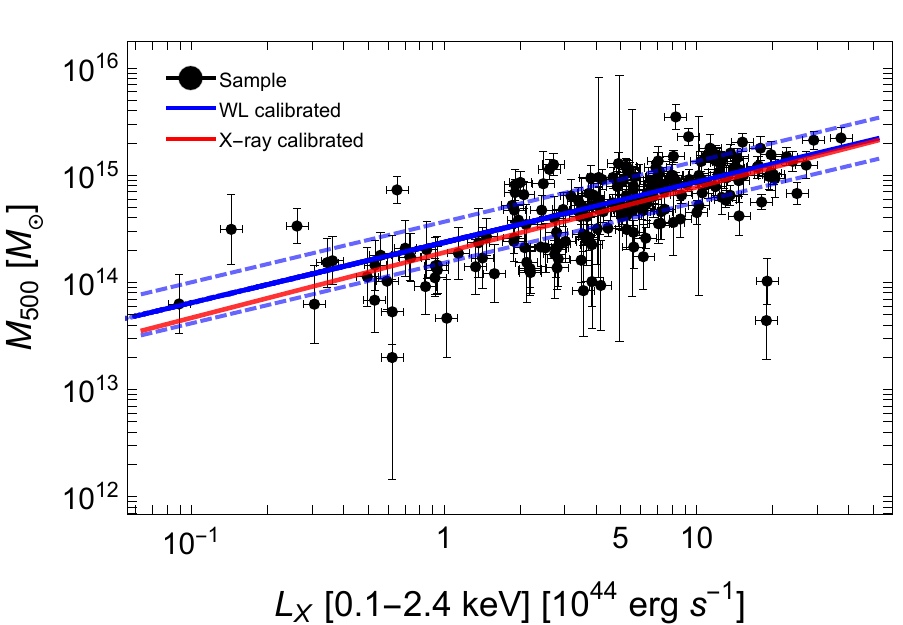}}
       \caption{Soft band luminosity vs mass $M_{500}$. The black points with error bars picture the calibration sample, $M_\text{LC2}$. The blue points picture the clusters with predicted masses  $M_\text{WLc}$ at the position of the measured luminosity. The dashed blue lines show the median scaling relation (full blue line) plus or minus the intrinsic scatter. The red line tracks the scaling relation from \citet{arn+al10} used to calibrate X-ray masses in \citet{pif+al11}. Lines refer to the median redshift $z=0.122$. Masses are in units of $M_\odot$.}
	\label{fig_LX_M500}
\end{figure}

\begin{figure}
       \resizebox{\hsize}{!}{\includegraphics{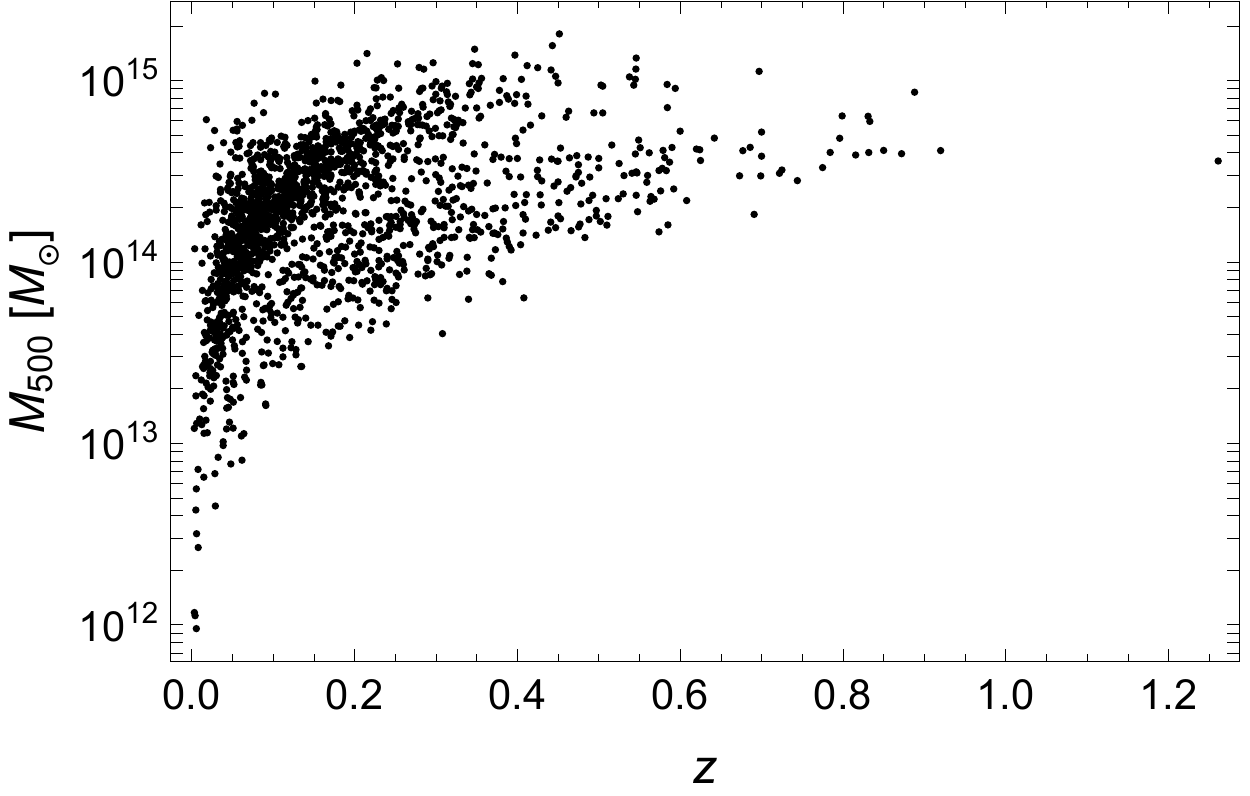}}
       \caption{Distribution of the MCXC clusters in the $M_{200}$-$z$ plane. Masses are WL calibrated and in units of $M_\odot$.}
	\label{fig_z_MWLc_MCXC}
\end{figure}

The MCXC \citep[Meta-Catalogue of X-ray detected Clusters of galaxies,][]{pif+al11} comprises 1743 unique X-ray clusters collected from available ROSAT All Sky Survey-based and serendipitous cluster catalogues. 

X-ray luminosities were systematically homogenized to the [0.1--2.4] keV band and standardized to an over-density of $\Delta=500$. Uncertainties are not provided. For our analysis, we fixed the statistical uncertainty on the luminosity to 10 per cent. As the LC$^2$, the MCXC is not statistically complete.

Luminosities within $r_{500}$ quoted in the MCXC catalog were mostly obtained by rescaling the total luminosity under the assumption of an universal density profile or they were computed iteratively from the available aperture luminosities by assuming the $L$-$M$ relation from \citet{arn+al10}. The same $L$-$M$ was also used to estimate $M_{500}$, which we refer to as $M_\text{X}$ in the following.

MCXC and LC$^2$-single share 196 clusters, which constitute our calibration subsample. We identified counterparts in the LC$^2$-single catalogue by matching pairs whose redshifts differ for less than $\Delta z =0.05$ and whose angular separation in the sky does not exceed 2 arcminutes. 

WL calibrated masses are shown in Fig.~\ref{fig_LX_M500}. The catalog of WL calibrated masses is released with the paper. An extract is presented in Table~\ref{tab_mcxc_MWLc}. The MCXC clusters covers a large mass range. WL calibrated masses go from $M_{500}\sim 2 \times 10^{12} M_\odot$  to $\sim 2 \times 10^{15} M_\odot$, with a typical value of  $M_{500}\sim 2.3 \times 10^{14} M_\odot$ and with a dispersion of $\sigma_M\sim 1.9\times 10^{14} M_\odot$. The catalog spans a large redshift interval from the local universe to $z\ls 1.3$, with a median redshift of $z=0.136$. 

Due to the heterogeneous nature of the MCXC catalog, we could not correct for any distant dependent Malmquist bias. The composite selection criteria made the proxy distribution smooth without truncation at the lower end. We modelled it as a simple redshift-evolving Gaussian distribution. In this simple framework, the WL calibrated masses and the measured luminosities align very well with the scaling relation, see Fig.~\ref{fig_LX_M500}. Since we could not fully correct for selection effects, some biases can persist, see Sec.~\ref{sec_MWLc_comp}.

The distribution of MCXC clusters in the mass-redshift ($M_{500}$-$z$) plane, using the WL calibrated mass proxy, is shown in Fig.~\ref{fig_z_MWLc_MCXC}. Notwithstanding the heterogeneous selection criteria used to compile the sample, the mass limit steadily increases with redshift. However, we can distinguish two populations with quite distinct mass limits. MCXC exploits both RASS-based and serendipitous surveys. At a given redshift, serendipitously discovered clusters are less luminous and massive than those from RASS-based catalogues because the deeper exposures allow lower flux limits to be adopted \citep{pif+al11}.

\subsection{WL vs. X-ray calibration}
\label{sec_planck_WL_vs_X}

\begin{figure}
       \resizebox{\hsize}{!}{\includegraphics{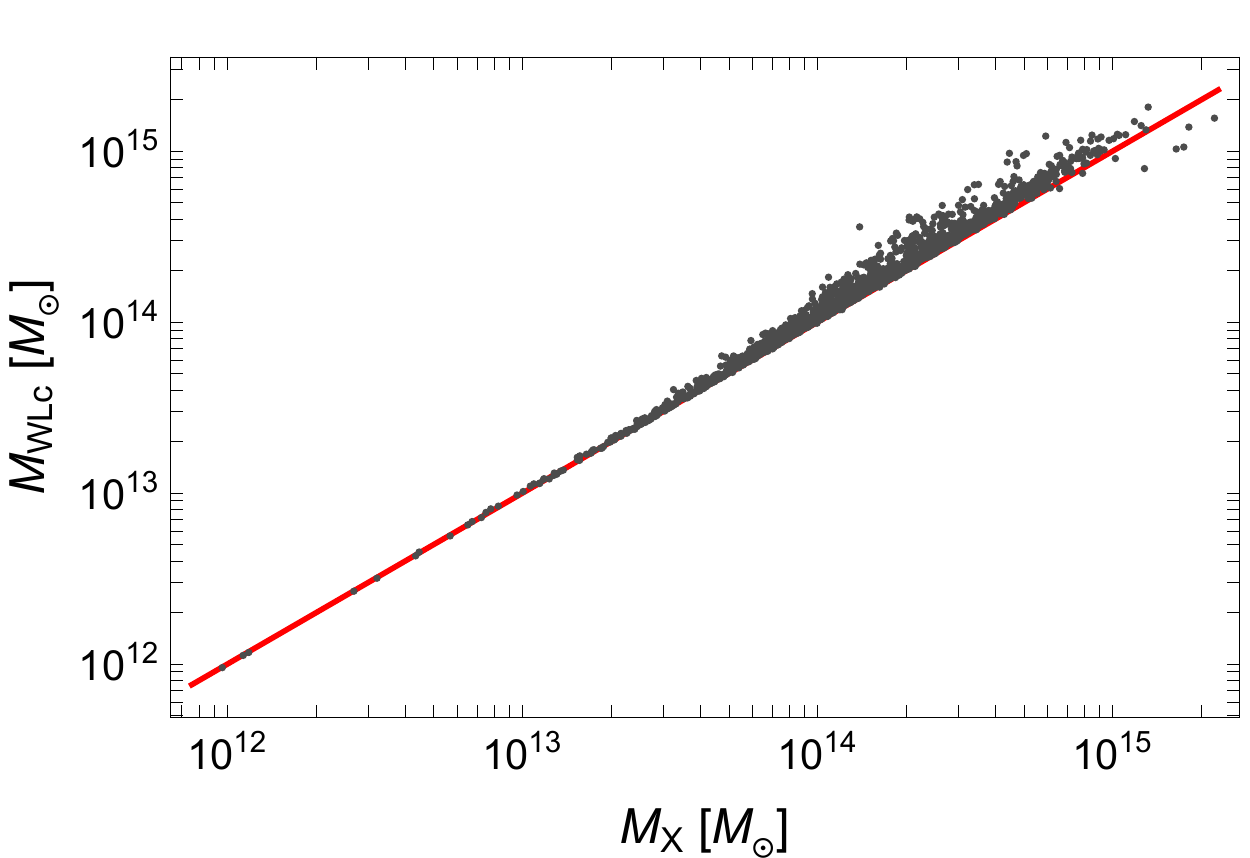}}
       \caption{MCXC masses, $M_\text{X}$,  versus WL calibrated masses, $M_\text{WLc}$. Masses are in units of $M_\odot$. The red line shows the bisection $M_\text{X}=M_\text{WLc}$.}
	\label{fig_MX_MWLc}
\end{figure}

\begin{figure}
       \resizebox{\hsize}{!}{\includegraphics{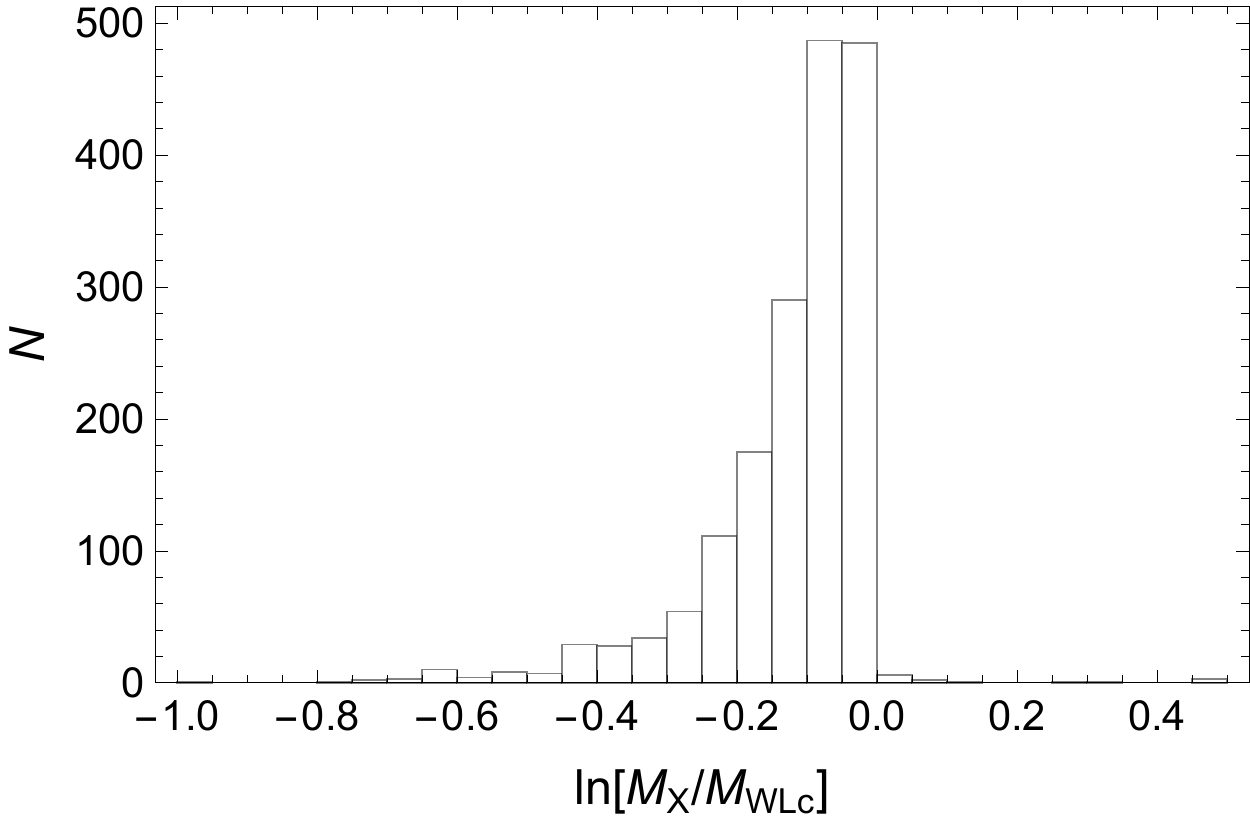}}
       \caption{Histogram of the logarithm of the ratio of the MCXC masses $M_\text{X}$ to the WL calibrated masses $M_\text{WLc}$.}
	\label{fig_MX_MWLc_histo}
\end{figure}

The WL calibrated masses differ from the X-ray masses which are underestimated on average by $\sim 14$ per cent, see Figs.~\ref{fig_MX_MWLc} and \ref{fig_MX_MWLc_histo}. In fact, our estimated scaling relation has a higher normalization than that used in \citet{pif+al11}, see Fig.~\ref{fig_LX_M500}. Most of the off-set can be attributed to the hydrostatic bias. Hydrostatic masses or X-ray calibrated masses can be underestimated by 20-30 per cent due to non-thermal sources of pressure in the gas, to unvirialized bulk motions and subsonic turbulence, to temperature inhomogeneity, and, to a lesser degree and mainly in the external regions, to the presence of clumps \citep{bat+al12,ras+al12,se+et15_comalit_I}. 

We can look for mass or redshift dependences of the bias. We proceeded as in Sec.~\ref{sec_planck_WL_vs_SZ} by comparing the WL to the X-ray calibrated masses of all the MCXC clusters. We fitted the relation
\beq
\ln \langle M_\text{X} \rangle =b_\text{X} + \beta_\text{X}  \ln \langle M_\text{WLc} \rangle +\gamma_\text{X} \ T .
\eeq
The X-ray calibrated masses in the MCXC catalog are provided without uncertainties. We adopted a statistical uncertainty of $\sim40$ per cent, which reflects the intrinsic scatter of the X-ray luminosities in the MCXC catalog \citepalias{se+et15_comalit_IV}.

The bias is nearly mass independent ($\beta_\text{X} =1.02\pm0.01$). In the local universe ($z=0.01$), the bias is negligible ($b_\text{X} =-0.02 \pm0.01$), but it becomes more prominent with redshift ($\gamma_\text{X} =-3.2\pm0.2$). 

This is the same trend found for SZ calibrated masses, see Sec.~\ref{sec_planck_WL_vs_SZ}. In fact, we are using the same calibration sample, LC$^2$-single. Any trend in the mass calibration can be only ascribed to redshift dependences.


\citet{sim+al17b} used stacked weak lensing measurements to calibrate the average masses for 166 luminous MCXC clusters. They found evidence that the $M_\text{X}$'s are approximately 15 to 30 per cent lower than the weak lensing masses over the range of masses probed in their  subsample (0.9 to 15$\times10^{14}M_\odot$), in agreement with our result.

\section{Catalog comparison}
\label{sec_MWLc_comp}

\begin{table}
\caption{Comparison of the WL calibrated masses within $r_{500}$ of the PSZ2, redMaPPer, and MCXC catalogs. We quote the mean $\ln$ differences in mass for sample pairs. Entries are in the format: $(N_\mathrm{cl}), \mu(\pm \delta\mu)\pm \sigma(\pm\delta \sigma)$, where $N_\mathrm{cl}$ is the number of clusters in common between the samples; $\mu$ is the central estimate of the difference in natural logarithm $\ln ( M_{500}^{\mathrm{row}}/M_{500}^\mathrm{col})$, with associated uncertainty $\delta \mu$; $\sigma$ is the dispersion with associated uncertainty $\delta \sigma$. $M_{500}^{\mathrm{row}}$ ($M_{500}^\mathrm{col}$) refers to the sample indicated in the corresponding row (column). Quoted values are the bi-weight estimators.}
\label{tab_comp_MWLc_500}
\centering
\begin{tabular}[c]{|ccccc}     
				&	PLANCK			&	MCXC	              	        \\ 
\hline
  				&	 	                         &  	$(370)$ 				 \\
PLANCK		        &	 ---	                         &  	$0.17(\pm0.01)$ 		 \\
  				&	 	                         &  	$\pm0.23(\pm0.01)$	         \\
\hline
 				& 	$(189)$  	                 & 	$(324)$ 			        \\
redMaPPer 	        & 	$-0.03(\pm0.02)$      & 	$0.24(\pm0.03)$ 	        \\
 				& 	$\pm0.30(\pm 0.02)$& 	$\pm0.46(\pm 0.02)$ 	 \\
\hline
	
\end{tabular}
\end{table}

\begin{figure}
\begin{tabular}{c}
\resizebox{\hsize}{!}{\includegraphics{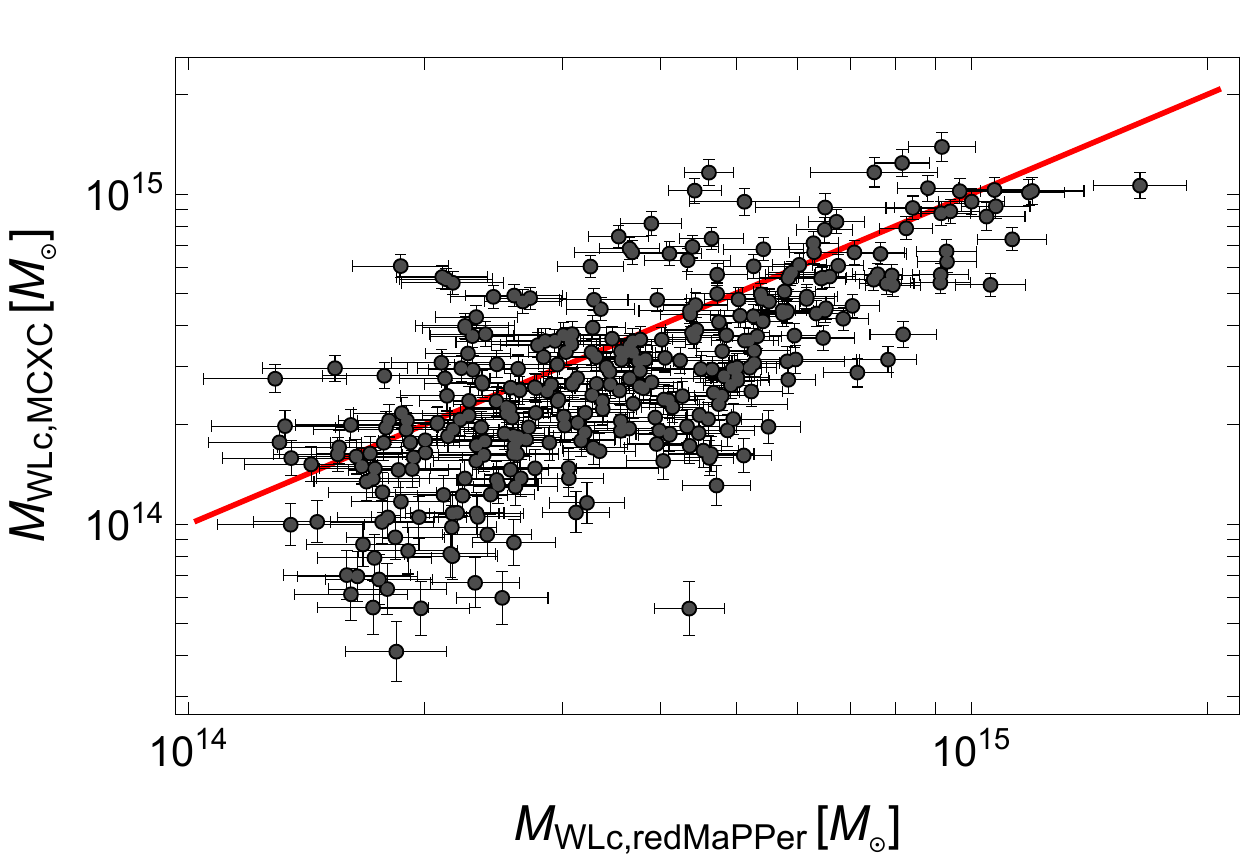}} \\
\noalign{\smallskip}  
\resizebox{\hsize}{!}{\includegraphics{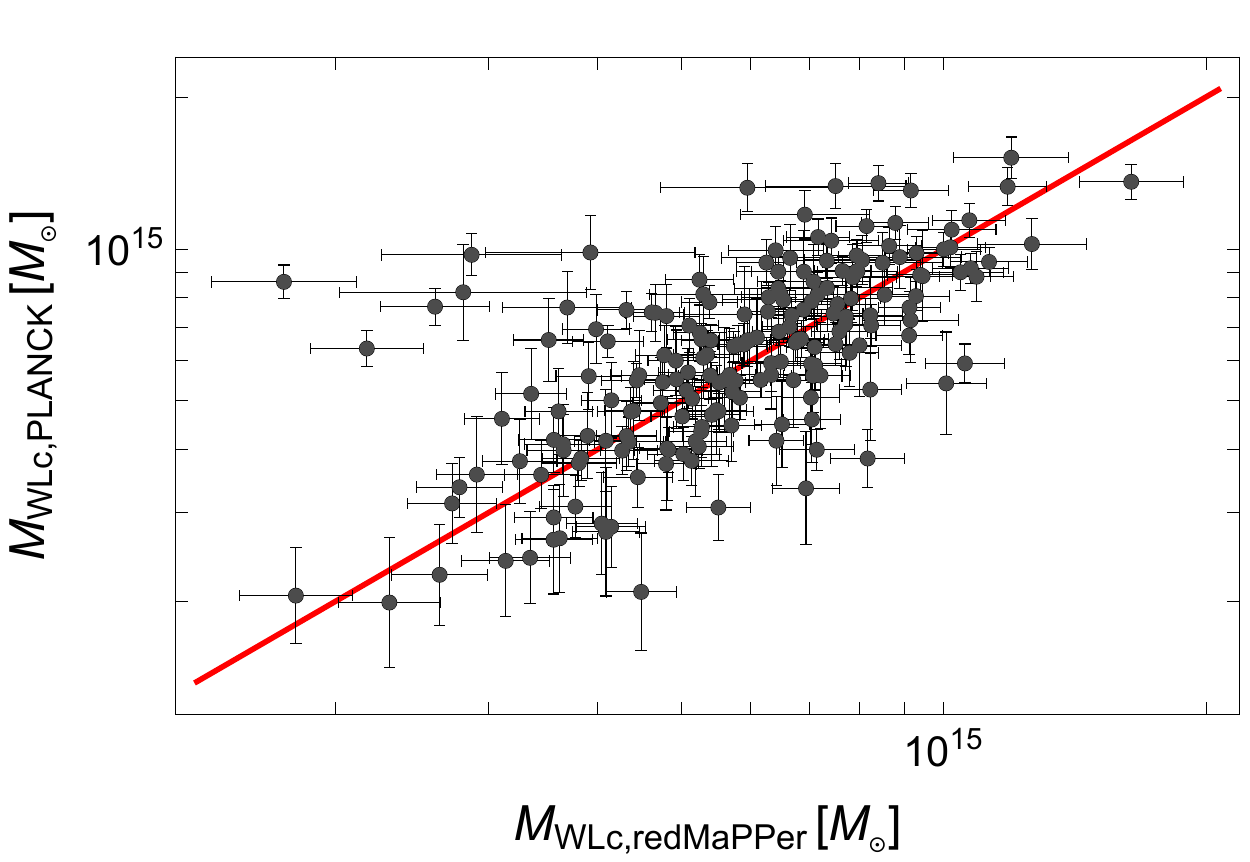}} \\
\noalign{\smallskip}  
\resizebox{\hsize}{!}{\includegraphics{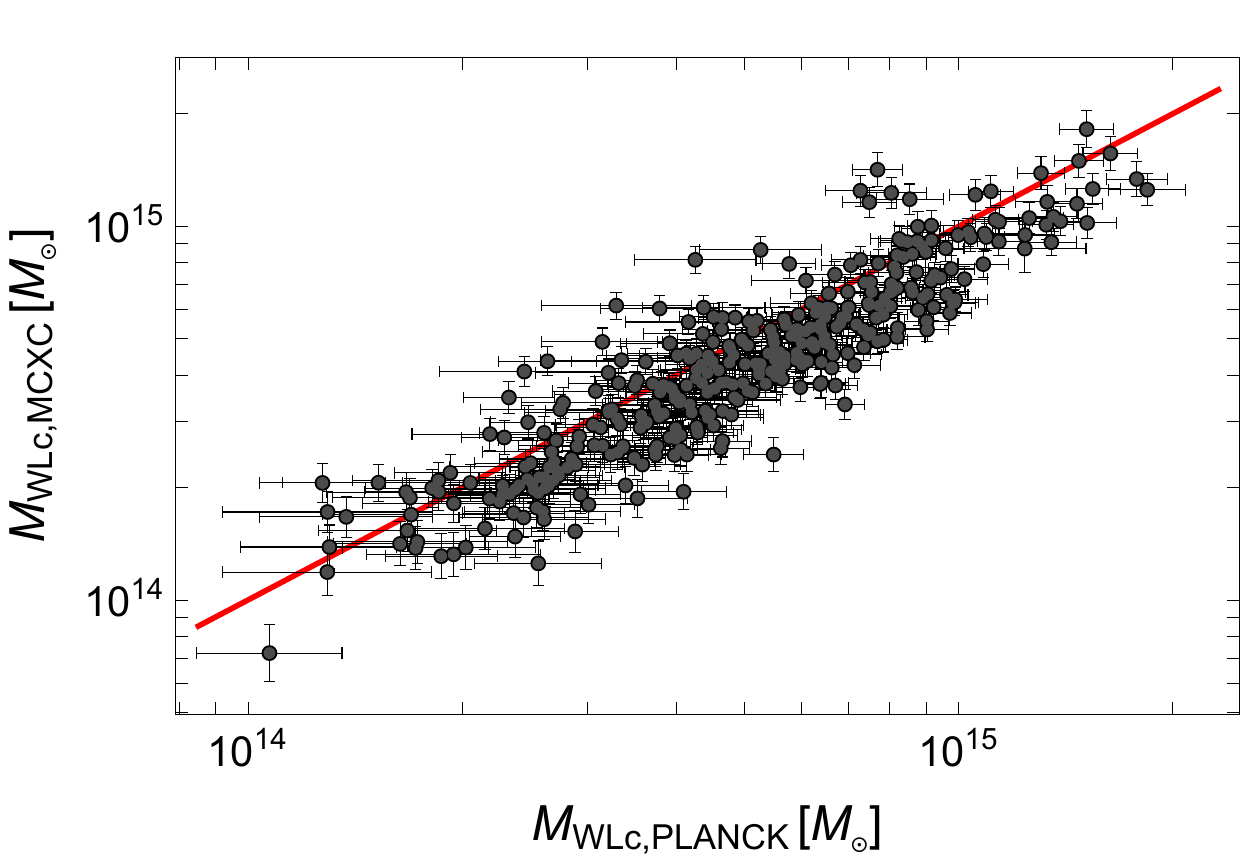}} \\
\end{tabular}
\caption{{\it Top panel}: WL calibrated mass of the redMaPPer clusters versus the masses of the MCXC clusters. {\it Middle panel}: WL calibrated masses of the redMaPPer clusters versus the masses of the {\it Planck} clusters. {\it Bottom panel}: WL calibrated masses of the PSZ2 clusters versus the masses of the MCXC clusters. The red lines show the bisectors. Masses are within $r_{500}$ and they are in units of $M_\odot$.}
\label{fig_MWLc_comparison}
\end{figure}

\begin{figure}
\begin{tabular}{c}
\resizebox{\hsize}{!}{\includegraphics{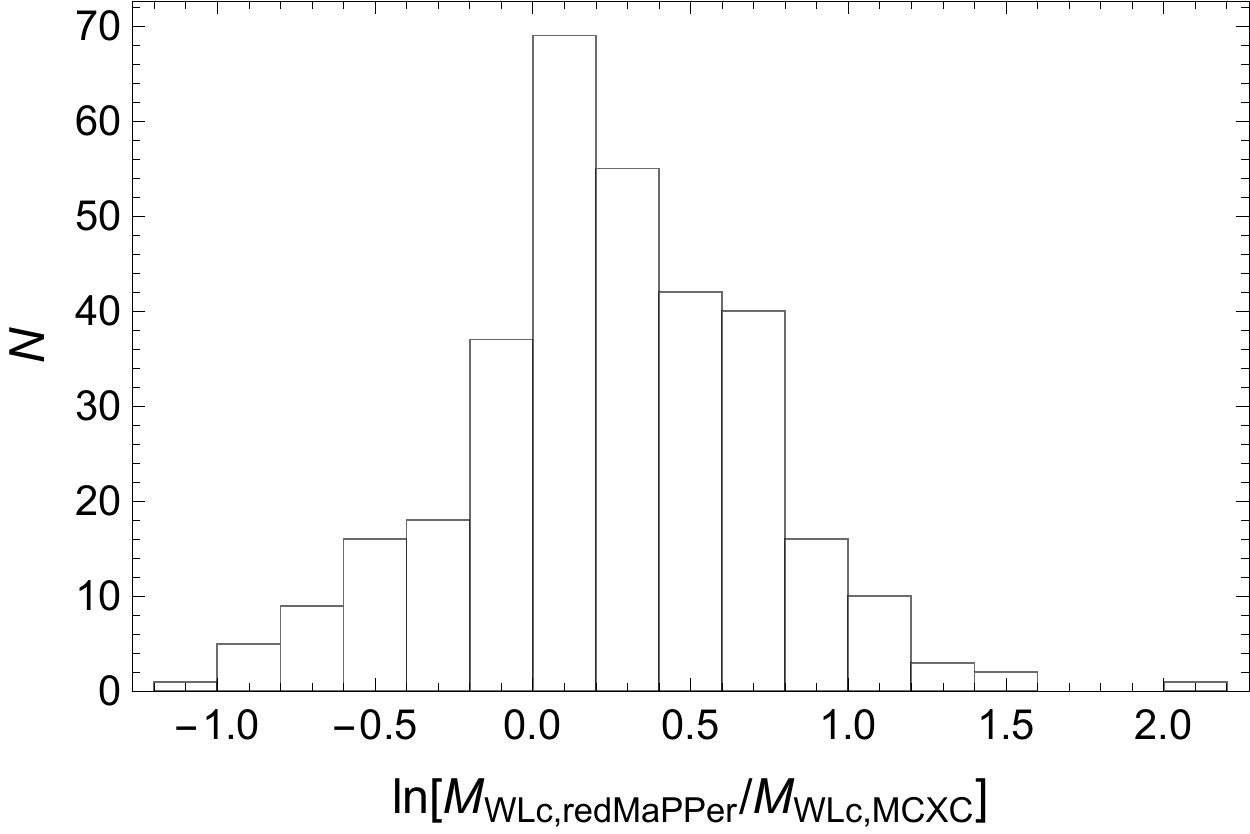}} \\
\noalign{\smallskip}  
\resizebox{\hsize}{!}{\includegraphics{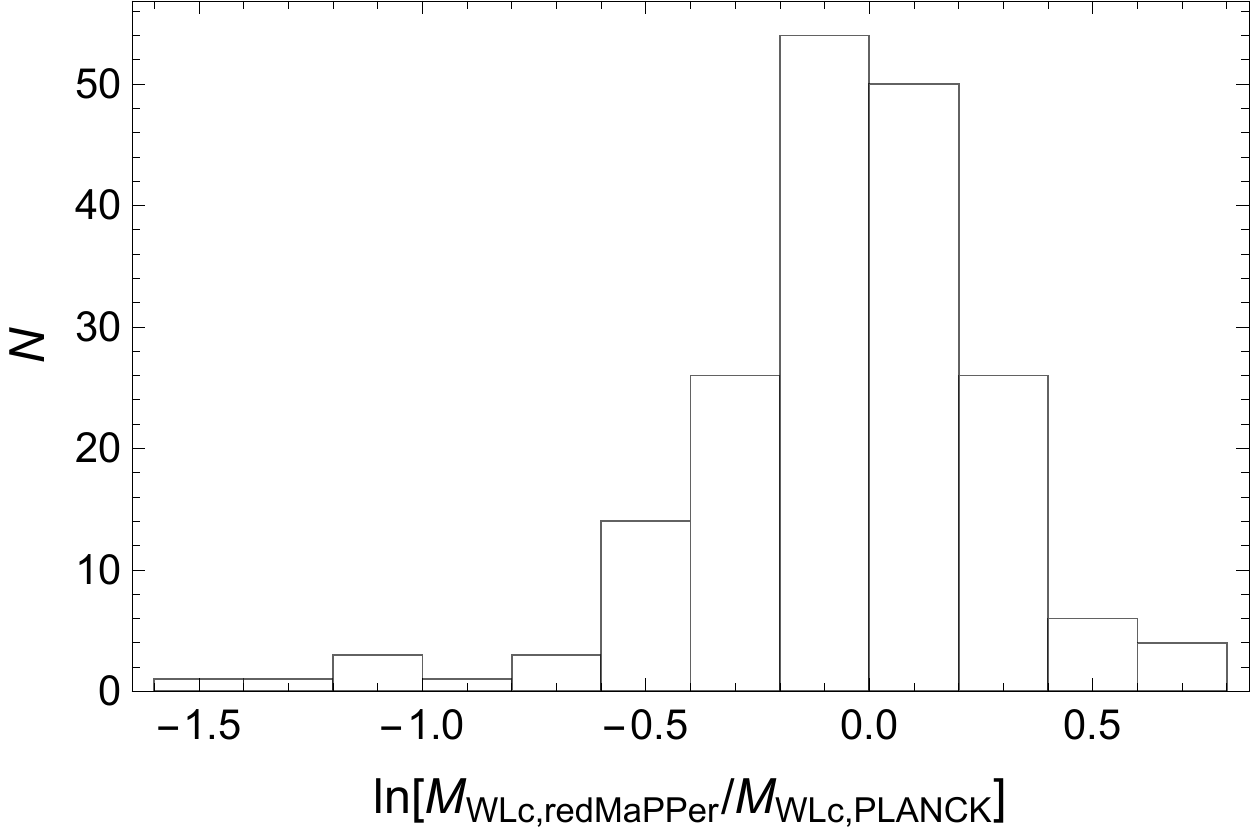}} \\
\noalign{\smallskip}  
\resizebox{\hsize}{!}{\includegraphics{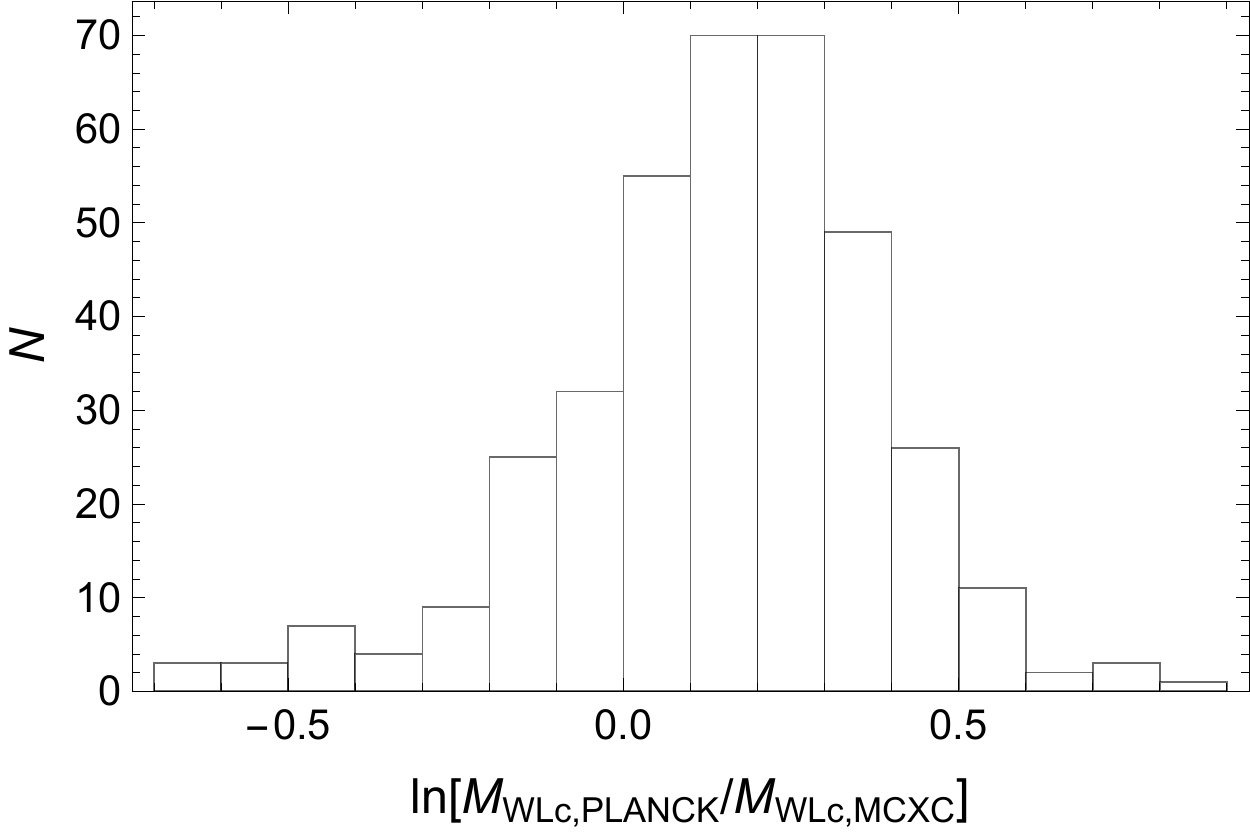}} \\
\end{tabular}
\caption{{\it Top panel}: Histogram of the natural logarithm of the ratio of the WLc masses of the redMaPPer clusters to the WLc masses of the MCXC clusters. 
{\it Middle panel}: Histogram of the natural logarithm of the ratio of the WLc masses of the redMaPPer clusters to the WLc masses of the PSZ2 clusters. {\it Bottom panel}: Histogram of the natural logarithm of the ratio of the WLc masses of the PSZ2 clusters to the WLc masses of the MCXC clusters.}
\label{fig_MWLc_comparison_histo}
\end{figure}

The PSZ2, the redMaPPer, and the MCXC samples have a certain degree of superposition, which allowed us to check for consistency in our WL calibrated masses. Results are summarized in Table~\ref{tab_comp_MWLc_500} and Figs.~\ref{fig_MWLc_comparison} and \ref{fig_MWLc_comparison_histo}.

We identified clusters in catalogue pairs by matching clusters whose redshifts differ for less than $\Delta z =0.05$ and whose angular separation in the sky does not exceed 2 arcminutes. 

WL calibrated masses for the redMaPPer clusters are within $r_{200}$. To compare  with {\it Planck} and MCXC clusters, whose masses are within $r_{500}$, we rescaled the redMaPPer masses assuming a Navarro-Frenk-White profile \citep{hu+kr03,ser15_comalit_III} with the mass-concentration relation derived in \citet{bha+al13}, which we rescaled to a value of $\sigma_8$, i.e. the amplitude of the matter power spectrum, convenient to the latest {\it Planck} cosmological analysis \citep{planck_2015_XIII}. We checked that the dependence of the rescaling on the adopted mass-concentration relation and cosmological parameters is negligible by also considering alternative derivations, e.g. \citet{du+ma14}.

The WL calibrated masses show good agreement, see Table~\ref{tab_comp_MWLc_500}. Discrepancies are statistical and well within the dispersion values. This is comfortable since we used the same calibration catalog of WL masses, i.e. LC$^2$-single.

The WL calibrated masses and the SZ masses of the {\it Planck} clusters, whose comparison is depicted in Figs.~\ref{fig_MSZ_MWLc} and \ref{fig_MSZ_MWLc_histo}, relied on the same proxy measurements and were then correlated. The same consideration holds for the WL or X-ray calibrated masses of the MCXC clusters, see Figs.~\ref{fig_MX_MWLc} and \ref{fig_MX_MWLc_histo}. This correlation strongly reduces the scatter. On the other hand, WL calibrated masses of PSZ2, redMaPPer, and MCXC clusters rely on independent proxies. This explains the quite large scatter in the mass comparison which mostly comes from the combined intrinsic scatters of the scaling relations. However, a good degree of correlation still persists, since the three catalogs of WLc masses exploited the same calibration catalog, i.e. LC$^2$-single. This prevented us from combining the WLc masses a posteriori.

Whereas forecasted masses are in agreement and differences are smaller than the statistical scatter, we found a tentative trend for the WL calibrated masses of the MCXC clusters to be over-estimated in comparison to either the redMaPPer or the PSZ2-based estimates. This can be ascribed to some residual selection effect. Most clusters in the MCXC sample were flux-selected but we could not correct for the Malmquist bias due to the unknown observational thresholds, which are not provided in the catalog. The X-ray luminosities of the clusters near the threshold are likely over-estimated, which on turn over-estimate the mass predictions.

\section{Reliability of the calibration sample}
\label{sec_reli}

\begin{table}
\caption{Comparison of WL masses within $r_{500}$ from independent analyses. Entries are as in Table~\ref{tab_comp_MWLc_500}.}
\label{tab_comp_MWL_500}
\resizebox{\hsize}{!} {
\begin{tabular}[c]{|cccccc}     
				&	CLASH			&	LoCuSS	                         &	WtG			        \\ 
\hline
  				&	$(6)$ 			&  	$(21)$ 				&	$(17)$ 			 \\
CCCP 			&	$-0.58(\pm0.22)$	&  	$0.02(\pm0.09)$ 		&	$-0.22(\pm0.07)$ 	 \\
  				&	$\pm0.28(\pm0.20)$	&  	$\pm0.39(\pm0.09)$	        &	$\pm0.25(\pm0.06)$  \\
\hline
 				& 	  			        & 	$(6)$ 				& 	$(16)$  		        \\
CLASH 		        & 	---	                        & 	$0.13(\pm0.09)$ 		& 	$0.07(\pm0.13)$ 	 \\
 				& 	                                & 	$\pm0.13(\pm 0.10)$ 	& 	$\pm0.40(\pm0.07)$	 \\
\hline
				&		 			&	                                         &	$(15)$			 \\
LoCuSS		        &	---			 	&	---                       		&	$-0.16(\pm0.12)$ 	 \\
				&					&	                                         &	$\pm0.37(\pm0.07)$	 \\
\hline
	
\end{tabular}
	}
\end{table}

\begin{table}
\caption{Comparison of WL masses within 1~Mpc from independent analyses. The notation follows Table~\ref{tab_comp_MWL_500}.}
\label{tab_comp_MWL_1Mpc}
\resizebox{\hsize}{!} {
\begin{tabular}[c]{|cccccc}     
				&	CLASH			&	LoCuSS	                         &	WtG			        \\ 
\hline
  				&	$(6)$ 			&  	$(21)$ 				&	$(17)$ 			 \\
CCCP 			&	$-0.42(\pm0.02)$	&	$0.02(\pm0.04)$	        &	$-0.15(\pm0.04)$ 	 \\
  				&	$\pm0.03(\pm0.03)$	&	$\pm0.22(\pm0.07)$  	&	$\pm0.16(\pm0.04)$  \\
\hline
 				& 	  			        & 	$(6)$ 				& 	$(16)$  		        \\
CLASH 		        & 	---	                        & 	$0.12(\pm0.05)$ 		& 	$0.03(\pm0.08)$ 	 \\
 				& 	                                & 	$\pm0.09(\pm 0.08)$ 	& 	$\pm0.26(\pm0.05)$	 \\
\hline
				&		 			&	                                         &	$(15)$			 \\
LoCuSS		        &	---			 	&	---                       		&	$-0.11(\pm0.08)$ 	 \\
				&					&	                                         &	$\pm0.25(\pm0.05)$	 \\
\hline
	
\end{tabular}
	}
\end{table}

\begin{table}
\caption{Homogeneous subsamples in the WL calibration sample used for the {\it Planck} clusters. We list the subsamples in LC$^2$-single with at least 6 {\it Planck} matches. Col.~1: sample name. Col.~2: number of clusters, $N_\mathrm{cl}$, in LC$^2$-single with a {\it Planck} counterpart. Cols.~3 and 4: typical redshift and dispersion. Cols.~5 and 6: typical mass and dispersion. Typical values and dispersions are computed as bi-weighted estimators. Masses are in units of $10^{14}M_\odot$.}
\label{tab_LC2_subsamples}
\resizebox{\hsize}{!} {
\begin{tabular}{ l r c c r c } 
 Sample		 &	$N_\mathrm{Cl}$	&	$z$	&	$\sigma_z$	&	$M_{500}$	&	$\sigma_{M_{500}}$  \\ 
\hline
WtG &  26  &  0.34 &  0.12 &  12.1 &  5.3 \\
LoCuSS     &  22  &  0.22 &    0.04 &  6.8&  2.5   \\
\citet{cyp+al04}  &  18 &  0.11 &   0.04  &  3.9  &  2.2 \\
CLASH   &  13  &  0.37 &    0.13 &  11.3 &  3.3  \\
CCCP  &  11  &  0.22 &    0.06 &  8.6 &  2.4 \\
\citet{gru+al14}     &  7   &  0.29 &    0.12 &  4.2 &  2.1 \\
\citet{ogu+al12}      &  6   &  0.47 &    0.10 &  6.3&  3.2   \\
\hline
	
\end{tabular}
	}
\end{table}

\begin{figure}
       \resizebox{\hsize}{!}{\includegraphics{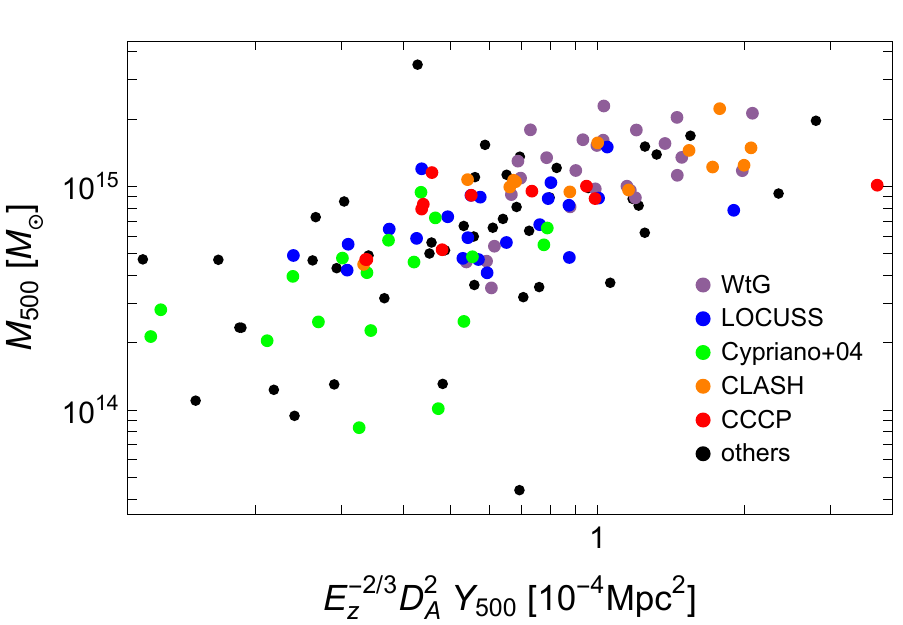}}
       \caption{The composition of the LC$^2$-single sample with PSZ2 counterparts. Masses of homogeneous subsamples with at least 10 matches with the {\it Planck} catalog are plotted versus the SZ flux.}
	\label{fig_YSZ_M500_LC2}
\end{figure}

At the present time, there are three main reasons to prefer composite, heterogeneous WL compilations to homogeneous calibration catalogs. Firstly, the significantly improved statistics. The largest available WL samples with homogeneous data-analysis and well measured masses consist of nearly fifty clusters, see Table~\ref{tab_samples}, an order of magnitude smaller than LC$^2$-single.

Secondly, current WL mass determinations by competing groups are not consistent \citepalias{se+et15_comalit_I}. Quoted statistical uncertainties of $\sim$10-20 per cent are smaller than the differences between independent mass determinations, which can be as large as $\sim 50$ per cent, see Tables~\ref{tab_comp_MWL_500} and \ref{tab_comp_MWL_1Mpc}. This discrepancy hints at systematics effects still to be fully understood. In presence of slightly compatible or discrepant analyses, it can be safer to consider all the results rather than to arbitrarily prefer a single data-set \citep{got+al01}.

Thirdly, for clusters with multiple analyses LC$^2$-single promotes the WL study exploiting the most deep observations and the most complete multi-band coverage. Analyses based on the better data are hopefully the more reliable and systematics free. Further considerations on heterogeneous samples can be found in \citet{pif+al11} and \citetalias{ser15_comalit_III}.

Ideally, one could introduce parameters which quantify the systematic off-sets of each sub-catalog relative to systematics-free measurements and associate priors to them. The systematic off-sets could then be solved simultaneously with the scaling relation. We do not pursue this full Bayesian treatment but in the following we check for systematics related to heterogeneity and the unknown selection function.

\subsection{Heterogeneity}
\label{sec_hete}

The above considerations are strong but general. They might fail in specific contexts. To further assess advantages, problematics and systematics related to heterogeneous calibration samples we focus in the following on the {\it Planck} clusters. The same kind of considerations hold for the redMaPPer and MCXC clusters, whose calibration WL samples have similar size and composition.

As shown in \citetalias{ser+al15_comalit_II}, the effect of the mass calibration is dramatic when considering the bias of {\it Planck} masses. The preference for a calibration sample drives the estimate of the bias. We repeated the analysis of Sec.~\ref{sec_planck_bias} for some homogeneous samples. Results are summarized in Table~\ref{tab_bias_MSZ}. Estimated biases range from $b_\text{SZ}\sim -0.2$ to $-0.4$. This is reflective of the inconsistent WL mass calibrations, which pushed us to consider LC$^2$-single as the reference sample to minimise systematics. We remark that different samples probe different mass and redshift ranges. The measured bias has to be intended as a sample-weighted mean. However, mass and redshift differences in the various samples are not so large to explain the bias discrepancy, which is linked to inconsistent mass measurements.

The subsample of LC$^2$-single with {\it Planck} counterparts used for calibration relies on 29 independent WL analyses, with only 7 subsamples being represented by more than six clusters, see Table~\ref{tab_LC2_subsamples}. There is substantial overlap among the different subsamples in both mass and redshift ranges, see Table~\ref{tab_LC2_subsamples} and Fig.~\ref{fig_YSZ_M500_LC2}. This guarantees that any systematic error affecting single subsamples is compensated for, even though at the expense of an increased scatter.

To further assess the stability of the heterogeneous LC$^2$-single as a calibration sample, let us consider a rather extreme pathological case, when the most represented subsample, i.e. the WtG subsample, has mass estimates systematically over- or under-estimated. The WtG clusters sample the high-mass end of the cluster distribution and are expected to have significant leverage in the determination of the scaling relation. Notwithstanding these extreme circumstances, even the effect of a very significant systematic bias of $\pm 10$ percent would impact only a fraction of the full sample (the WtG subsample constitutes $\sim 20$ per cent  of the total calibration sample) and we expect an overall systematic mass shift in the calibration sample of the order of $\sim \pm 2$ per cent. 

We repeated the analysis of Sec.~\ref{sec_planck_WL} assuming a systematic error in the WtG subsamples. A $+(-)10$ per cent systematic error produces an over- (under-)estimation of the WL calibrated masses of $\sim$1.7, 0.3, and 0.5 per cent for the clusters in the LC$^2$ catalog, for the cluster with no WL counterpart, and for the full {\it Planck} sample, respectively. As expected, even a large systematic effect is compensated by the composite nature of the calibration sample. It shows up mainly for clusters in the calibration samples, whereas its effect is very negligible for the clusters with no WL data.

The effect of a systematic error is further reduced by the Bayesian analysis. The WL calibrated masses are seen as unknown variables in the regression. Our employed Bayesian approach assumes that true masses are well behaved and align well along the scaling relation. Significant deviations off the scaling by some clusters, which characterise clusters affected by strong systematics, are penalized in the hierarchical model.

\subsection{Unknown selection function}

\begin{figure}
\begin{tabular}{c}
\resizebox{\hsize}{!}{\includegraphics{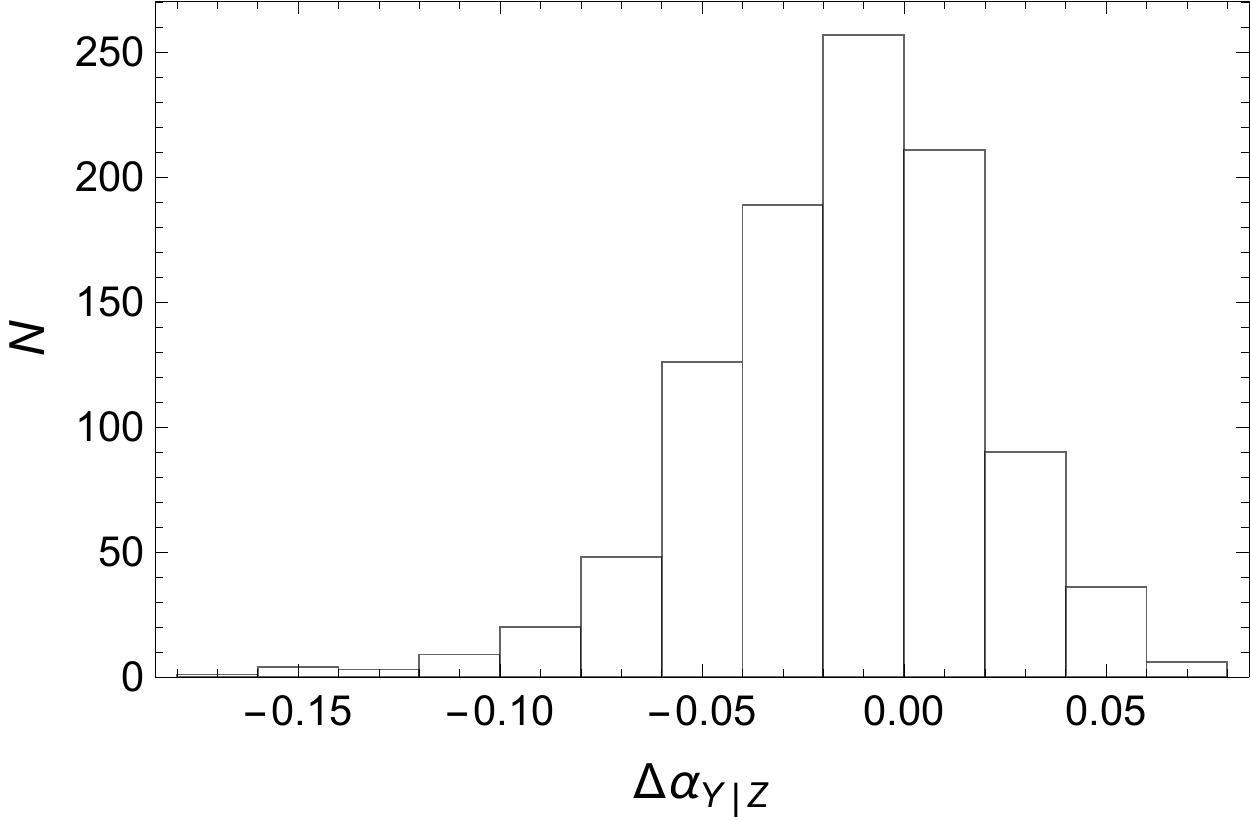}} \\
\noalign{\smallskip}  
\resizebox{\hsize}{!}{\includegraphics{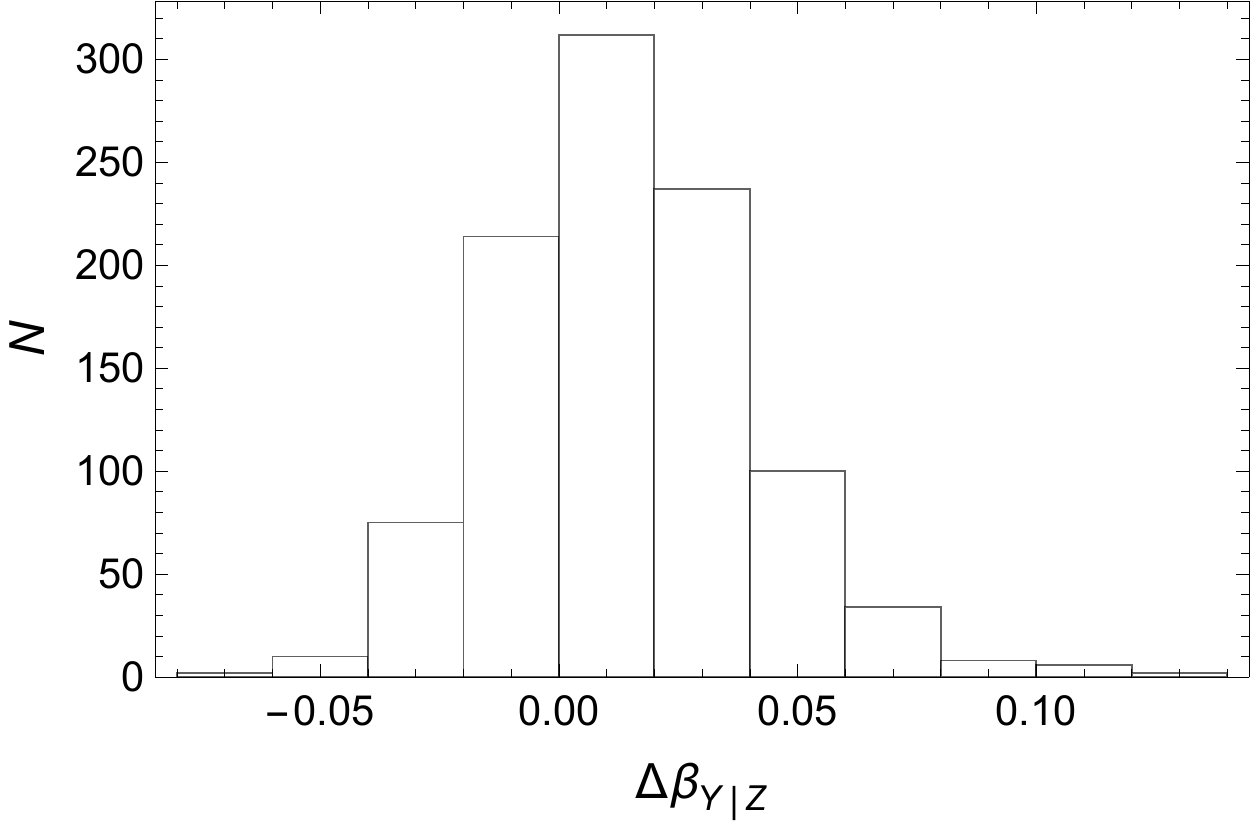}} 
\end{tabular}
\caption{{\it Top panel}: Histogram of the differences between the measured normalisations and the true one for the $M_\text{WL}-Y_\text{ZS}$ simulated samples selected in X-ray luminosity. {\it Bottom panel}: Same as above but for the slope.}
\label{fig_histo_sim_Delta_alpha_beta}
\end{figure}

\begin{figure}
\begin{tabular}{c}
\resizebox{\hsize}{!}{\includegraphics{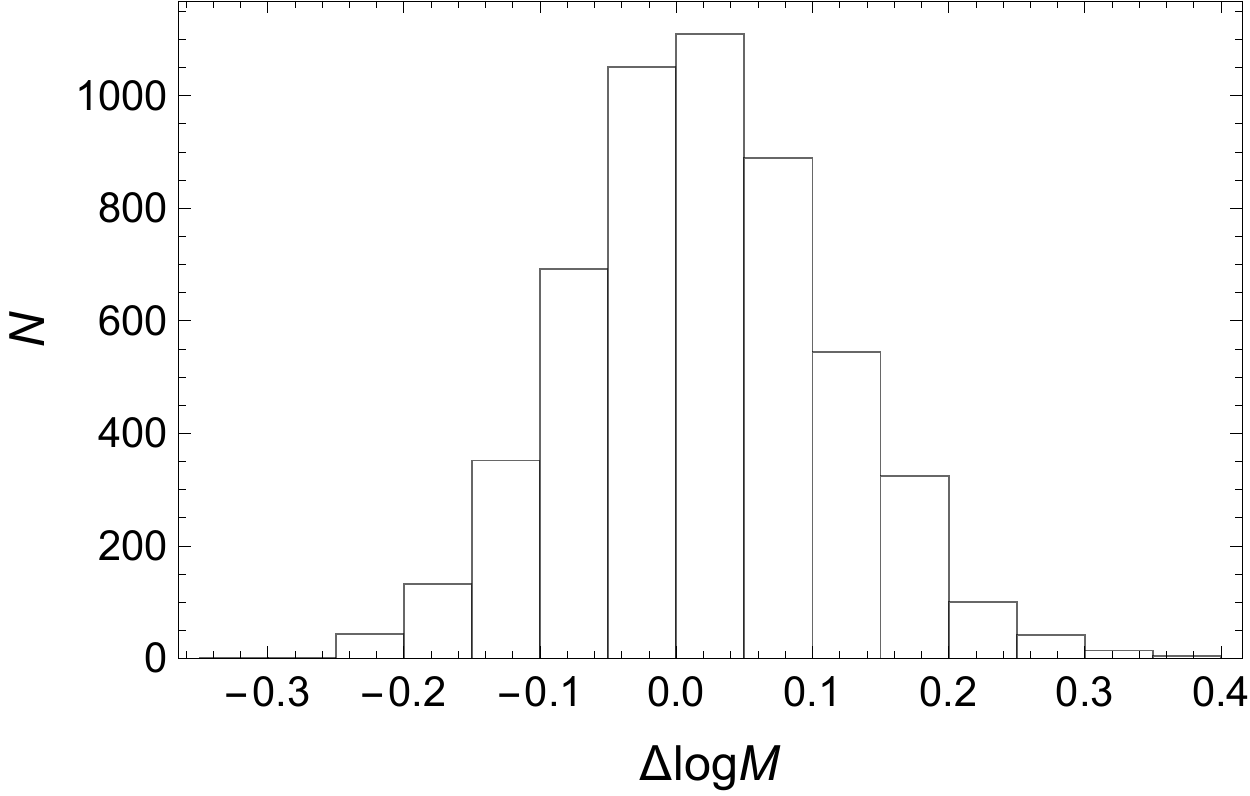}} 
\end{tabular}
\caption{Histogram of the differences between the measured masses and the true one for the $M_\text{WL}-Y_\text{ZS}$ simulated samples selected in X-ray luminosity.}
\label{fig_histo_sim_Delta_M}
\end{figure}

Most sources of scatter affect primarily one specific observable. However some of them can affect several simultaneously \citep{ang+al12,sta+al10}. Common sensitivities to the internal structure, orientation and environment of clusters, as well as to line-of-sight superposition of uncorrelated structure, can make scatters between different mass-observable relations positively correlated  \citep{ang+al12}. The degree of this correlation is still to be firmly established. Results from numerical simulations strongly depend on adopted schemes and specific physical treatments.

Scatter covariance might affect our results since the selection function of the calibration sample is unknown. Let us make the hypothetical case that the calibration sample preferentially includes bright X-ray clusters. As a result of covariance between WL mass, SZ flux and X-ray luminosity, at fixed $Y_\text{SZ}$ the calibration sample will preferentially include more massive (and less scattered) clusters than the entire {\it Planck} population. This will bias the measured scaling relation between mass and Compton parameter and the forecasted masses of the remaining population. Similar considerations might apply to richness and redMaPPer clusters.

Some arguments can be formulated to demonstrate that these biases should be negligible in our analysis. Firstly, the LC$^2$ is an unfiltered collection of all the clusters and groups with measured WL mass \citepalias{ser15_comalit_III}. The included clusters were originally detected within either X-ray, optical, SZ, or shear surveys. Serendipitous detections, follow-ups and targeted observations are included too. The WL signal is low and LC$^2$ mostly samples the very massive end of the halo mass function. 
Large and heterogeneous samples may reduce the effect of selection biases \citepalias{ser15_comalit_III}.
As far as mass and redshift are considered, the LC$^2$ sample is an unbiased sub-sample of the {\it Planck} clusters with very high signal to noise \citepalias{ser+al15_comalit_II}. The price to pay for heterogeneity is the increased scatter.

Secondly, the unbiased nature of our calibration subsamples is further indicated by the analysis of the completeness. In the CoMaLit approach, the completeness function can be estimated from the data together with the scaling relation \citepalias{se+et15_comalit_IV}. In fact, the completeness functions of the {\it Planck} and redMaPPer clusters in LC$^2$-single follow those of the parent populations, see the upper panels in figures 6 and 8 and related discussion in \citetalias{se+et15_comalit_IV}.

Finally, for a more quantitative assessment of the bias related to the selection function of the calibration subsample, we tested our forecasting procedure on a toy-population of clusters. We simulated the pathologic case of X-ray luminous clusters preferentially included in the calibration subsample. In the following, we consider the case of the $M_\text{WL}-Y_\text{SZ}$ relation but similar results apply to the richness. 

Here, we use arbitrary units such that the cluster with unitary mass has unitary SZ flux and X-ray luminosity too. We simulated a log-normal parent population of masses with $\langle \log M \rangle = 0$ and $\sigma_{\log M}=0.5$. All simulated haloes lie at the same redshift. The observables are self-similarly related to the true mass,
\begin{eqnarray}
M_\text{WL} &=  & M\ , \\
Y_\text{SZ} & = & M^{5/3}\ , \\
L_\text{X} & =  & M^{4/3}\ .
\end{eqnarray}
In the above equations, we neglected the redshift-dependence since the redshift is constant. For the aims of this testing, we do not have to specify the over-density radius.

Observed values of WL mass, Compton integrated parameter and X-ray luminosity were simulated by scattering the self-similar values. We considered intrinsic scatters of $\sigma_{\log M_\text{WL}|\log M}=0.1$, $\sigma_{\log Y_\text{SZ}|\log M}=0.1$, and $\sigma_{\log L_\text{X} |\log M}=0.15$ in line with results from \citetalias{se+et15_comalit_IV}. Scatter values are strictly related to the measurement methodology and to the sample properties. Here, we just need approximated values. Conditional scatters were correlated through the factors $\rho_{\log M_\text{WL}\log Y_\text{SZ}}=0.6$, $\rho_{\log M_\text{WL}\log L_\text{X}}=0.4$ and $\rho_{\log Y_\text{SZ} \log L_\text{X}}=0.5$ in line with the results of \citet{ang+al12} based on the Millennium-XXL simulation. We were interested in systematic effects and we neglected measurement uncertainties.

To test our regression method, we selected clusters in X-ray luminosity and than we estimated the $M_\text{WL}$-$Y_\text{SZ}$ relation without using the information on the selection and on the scatter correlation. A calibration sample of 100 objects was randomly drawn from the parent population of clusters with $\log L_X > \sigma_{\log L_\text{X}}=(4/3)\sigma_{\log M}$. In analogy to our real cases we limited the calibration sample to the bright end. 

As a first check, we recovered the self-similar relation, $M = Y_\text{SZ}^{3/5}$, by regressing the SZ fluxes versus the WL masses of the only clusters in the calibration sample. In the regression, we considered conditional scatters with respect to the true mass for both $M_\text{WL}$ and $Y_\text{SZ}$, but we neglected the scatter correlation. We used a smoothly truncated Gaussian distribution to model the distribution of the unscattered values of the SZ fluxes. We did not model the Malmquist bias related to the selection in X-ray luminosity

This simplified regression scheme is accurate enough to recover the true scaling relation. In Fig.~\ref{fig_histo_sim_Delta_alpha_beta} we show the deviations of the normalizations and the slopes from the true values for one thousand simulated samples. 

Most of the selection effects are accounted for with a proper modelling of the covariate distribution. Since the sample size of 100 objects is quite small, we cannot constrain the scatter correlation from the data. However, the wrongly assumed values of $\rho_{\log M_\text{WL}\log Y_\text{SZ}}=0.0$ has a negligible effect. Notwithstanding the systematic errors in the modelling, the self-similar relation of the full population was well recovered. 

As a second check, we considered the effect of scatter covariance in forecasting. The calibration sample was selected as before, whereas the 530 masses to be forecasted were drawn from the population of low luminosity clusters, i.e. $\log L_X < (4/3)\sigma_{\log M}$. We neglected the conditional scatter in the covariate variable and we modelled the covariate distribution, including both the calibration subsample and the clusters with unknown masses, as a Gaussian. We simulated 10 samples to estimate the systematic errors in mass forecasting associated to the unknown selection function of the calibration sample and to the unknown (and not modelled) scatter covariance. Differences between true and recovered masses are shown in Fig.~\ref{fig_histo_sim_Delta_M}. As for the scaling relation, masses are well recovered. 

To the present level of accuracy, the systematic biases discussed in this section are small with respect to the statistical uncertainties and can be neglected. The main limitation brought by the heterogeneous calibration sample is the increased conditional scatter of the WL masses with respect to the true masses and the related loss of accuracy in mass forecasting.

\section{Conclusions}
\label{sec_conc}

Mass prediction given a proxy can be efficiently performed in the context of Bayesian hierarchical modelling. The key to efficiency is a proper treatment of measurement errors, Malmquist/Eddington biases, selection effects and redshift evolution.

Our proposed statistical method performs mass forecasting together with the determination of the scaling relation. Mass calibration relies on a reference subsample of clusters with measured masses. This subsample does not have to be representative of the full sample, whose distribution is modelled in the regression. However, we still assume that the scaling relation and the conditional scatter of the subsample can characterise the full sample. Thanks to the simultaneous determination of masses, scaling relation and proxy distribution, systematics are strongly reduced and statistical errors are properly assessed.

We applied the CoMaLit mass forecasting to three widely used cluster samples: the PSZ2 catalogue of {\it Planck} detected clusters, the redMaPPer catalogue of red-sequence filtered clusters, and the meta-catalogue of X-ray detected clusters MCXC. As calibration sample, we considered the subsamples of clusters with measured WL mass. Catalogues of weak lensing calibrated masses are released with the paper. The Bayesian analysis was performed with the publicly available \textsc{R}-package \texttt{LIRA}.

The considered samples were not selected by their WL properties. The PSZ2 and the redMaPPer catalogues were selected by SZ flux and optical richness, respectively. The MCXC clusters were selected based on their X-ray properties. We considered the full catalogues and we treated the masses of clusters without WL estimates as missing data. As far as the calibration subsample with measured WL masses follows the scaling relation of the full sample and it is affected by the same intrinsic scatter, there are no major selection effects related to the WL measurements. Accordingly, we modelled the distribution of the full sample, not the distribution of the subsample with measured WL mass.

{\it Planck}'s cluster count cosmology results \citep{planck_2015_XXIV} favour smaller values of the amplitude of the matter power spectrum, $\sigma_8$, and of the matter density parameter, $\Omega_\mathrm{M}$, than those from the {\it Planck}'s measurements of the primary CMB temperature anisotropies \citep{planck_2015_XIII}. The mass bias required to reconcile the two experiments is larger than current results and corresponds to $b_\text{SZ}=-0.42\pm0.04$ \citep{planck_2015_XXIV}. This assumes that the mass bias does not depend on cluster mass and redshift.

We found that whereas the average bias is $b_\text{SZ}\sim-0.25$, which it is not enough to explain the disagreement, it is strongly redshift dependent. SZ masses of very large clusters at high redshift, the objects which have the more weight in number count analyses, are strongly underestimated.

We provided catalogs of WL calibrated masses corrected for biases and selection effects which cover considerable mass and redshift ranges. We associated mass estimates to the redMaPPer clusters, which were not present in the original catalog, and we corrected the {\it Planck} or MCXC masses for the hydrostatic bias. The estimated error budget of the forecasted masses includes uncertainties associated to the proxy measurement, the calibration sample, the scaling relation and the intrinsic scatter, which contributes the most. Errors and biases due to data selection are corrected for. Still, the statistical error has not to be confused with the intrinsic scatter of the scaling relation. In the CoMaLit approach, we modelled the true masses as unscattered and perfectly fitting the scaling relation. It is the WL mass associated to the cluster which is a scattered realization of the true mass.

The next step in mass prediction is the application of scaling relations generalized to include the dependence upon two or more independent observables \citep{ett+al12}. \citet{ett13} showed that a generalized relation $M \propto L_\text{X}^\alpha M_\text{g}^\beta T_\text{X}^\gamma$, where $L_\text{X}$ and $T_\text{X}$ are the X-ray luminosity and temperature, respectively, and $M_\text{g}$ is the gas mass, can significantly reduce the intrinsic scatter on both galaxy group and cluster scales.

\section*{Acknowledgements}
M.S. thanks J.-B. Melin and E. Rozo for useful explanations. M.S. and S.E. acknowledge the financial contribution from contracts ASI-INAF I/009/10/0, PRIN-INAF 2012 `A unique dataset to address the most compelling open questions about X-Ray Galaxy Clusters', and PRIN-INAF 2014 `Glittering Kaleidoscopes in the sky: the multifaceted nature and role of galaxy clusters'. This research has made use of NASA's Astrophysics Data System (ADS) and of the NASA/IPAC Extragalactic Database (NED), which is operated by the Jet Propulsion Laboratory, California Institute of Technology, under contract with the National Aeronautics and Space Administration.

\bibliographystyle{mn2e_fix_Williams}

\setlength{\bibhang}{2.0em}

\setlength{\itemindent}{-2.5em}

\appendix

\section{Implementation}
\label{app_imple}

The mass forecasting was performed with the package \texttt{LIRA}. Let \texttt{x}, \texttt{y}, \texttt{delta.x}, \texttt{delta.y}, \texttt{covariance.xy}, \texttt{x.threshold}, and \texttt{z} be the vectors storing the values of $\mathbfit{x}$, $\mathbfit{y}$, $\mathbfit{$\delta_x$}$, $\mathbfit{$\delta_y$}$, $\mathbfit{$\delta_{xy}$}$, and $\mathbfit{z}$, respectively. If not stated otherwise, priors are set to the default distributions.

\begin{itemize}

\item For the mass forecasting of the {\it Planck} clusters in Sec.~\ref{sec_planck_WL}, we corrected for the distance dependent Malmquist bias by assigning the selection thresholds in the SZ signal, $\mathbfit{x}$$_\mathrm{th}$. No error was associated to the thresholds. The analysis was performed with the command

\noindent \texttt{> mcmc <- lira(x, y,~delta.x=delta.x, delta.y=delta.y, x.threshold=x.threshold,  z=z, distance='luminosity', gamma.sigma.Z.Fz='dt', Y.monitored=TRUE, X.monitored=TRUE, YZ.monitored=TRUE)},

\noindent where the covariate distribution is modelled as a redshift-evolving Gaussian function. On top of the default values, we also allowed for redshift evolution of the width of the distribution (\texttt{gamma.sigma.Z.Fz='dt'}).

Thanks to the options \texttt{YZ.monitored=TRUE}, \texttt{Y.monitored=TRUE} and \texttt{X.monitored=TRUE}, we can monitor the WLc calibrated masses, the WL masses and the covariate $X$, respectively.

\item The forecasting in Sec.~\ref{sec_redmapper} for the redMaPPer catalog was performed with

\noindent \texttt{> mcmc <- lira(x, y,~delta.x=delta.x, delta.y=delta.y, x.threshold=x.threshold,  z=z, distance='luminosity', gamma.sigma.Z.Fz='dt', mu.Z.min.0=x.min, sigma.Z.min.0='dunif(10\^~-6, 0.3)', Y.monitored=TRUE, X.monitored=TRUE, YZ.monitored=TRUE)}

\noindent  where we corrected for the distance dependent Malmquist bias by assigning the selection thresholds for each point (\texttt{x.threshold}). As covariate distribution, we used a redshift-evolving Gaussian function truncated at small values with a complementary error function centred on \texttt{mu.Z.min.0=x.min} and with dispersion \texttt{sigma.Z.min.0}. Here, the parameter \texttt{x.min} is fixed to the logarithm of the minimum observed richness.

\item The forecasting in Sec.~\ref{sec_mcxc} for the MCXC catalog was performed with

\noindent \texttt{> mcmc <- lira (x, y,~delta.x=delta.x, delta.y=delta.y,  z=z, distance='luminosity', gamma.sigma.Z.Fz='dt', Y.monitored=TRUE, X.monitored=TRUE, YZ.monitored=TRUE)}

\noindent where we could not correct for the distance dependent Malmquist bias and we modelled the covariate distribution as a redshift-evolving Gaussian function.

\end{itemize}

Further examples and catalogs can be found at \url{http://pico.bo.astro.it/\textasciitilde sereno/LIRA/}.

\end{document}